\documentstyle[12pt,amsmath,amssymb,graphicx,graphics,epsfig]{article}
\newcommand{\rr}{{\mathbb{R}}}
\newcommand{\R}{{\mathbb R}}
\newcommand{\calA}{{\cal A}}
\def\br{\begin{thebibliography}{abc}}
\def\er{\end{thebibliography}}

\def\beq{\begin{equation}}
\def\eeq{\end{equation}}
\def\bea{\begin{eqnarray}} 
\def\eea{\end{eqnarray}}
\def\nn{\nonumber}
\newcommand{\be}{\begin{equation}}
\newcommand{\ee}{\end{equation}}

\numberwithin{equation}{section}
\begin{document}

\begin{flushright}SU-4252-870 \end{flushright}
\vspace{0.2in}
\begin{center}{\Large\bf\boldmath
Quantum Fields on the Groenewold-Moyal Plane \footnotemark \\}
\end{center}
\footnotetext{Based on the lectures given by A. P. B. at the Workshop on Noncommutative Geometry, University of New Brunswick, Fredericton, Canada from the 21st to the 24th of June 2007.}
\bigskip
\centerline{\bf Earnest Akofor, A. P. Balachandran and Anosh Joseph}
\vspace{0.2cm}
\centerline{{\it Department of Physics, Syracuse University, Syracuse, New York 13210, USA}}
\vspace{0.2cm}
\centerline{E-mail: {\tt eakofor@phy.syr.edu}, {\tt bal@phy.syr.edu}, {\tt ajoseph@phy.syr.edu}}
\bigskip

\begin{abstract}

We give an introductory review of quantum physics on the noncommutative spacetime called the Groenewold-Moyal plane. Basic ideas like star products, twisted statistics, second quantized fields and discrete symmetries are discussed. We also outline some of the recent developments in these fields and mention where one can search for experimental signals.

\end{abstract}

\vfill

\newpage

\tableofcontents

\newpage

\renewcommand{\baselinestretch}{1}
\baselineskip 18pt

\section{Introduction}
Quantum electrodynamics is not free from divergences. The calculation of Feynman diagrams involves a cut-off $\Lambda$ on the momentum variables in the integrands. In this case, the theory will not see length scales smaller than $\Lambda^{-1}$. The theory fails to explain physics in the regions of spacetime volume less than $\Lambda^{-4}$. 

Heisenberg proposed in the 1930's that an effective cut-off can be introduced in quantum field theories by introducing an effective lattice structure for the underlying spacetime. A lattice structure of spacetime takes care of the divergences in quantum field theories, but a lattice breaks Lorentz invariance. 

Heisenberg's proposal to obtain an effective lattice structure was to make the spacetime noncommutative. The noncommutative spacetime structure is point-less on small length scales. Noncommuting spacetime coordinates introduce a fundamental length scale. This fundamental length  can be taken to be of the order of the Planck length. The notion of point below this length scale has no operational meaning. 

We can explain Heisenberg's ideas by recalling the quantization of a classical system. The point of departure from classical to quantum physics is the algebra of functions on the phase space. The classical phase space, a symplectic manifold $M$, consists of ``points" forming the pure states of the system. Every observable physical quantity on this manifold $M$ is specified by a function $f$. The Hamiltonian $H$ is a function on $M$, which measures energy. The evolution of $f$ on the manifold is specified by $H$ by the equation
\beq
\dot{f}=\{f, H\}
\eeq
where $\dot{f} = df/dt$ and $\{ \; ,\;  \}$ is the Poisson bracket.

The quantum phase space is a ``noncommutative space" where the algebra of functions is replaced by the algebra of linear operators. The algebra $ {\mathcal{F}}(T^*Q) $ of functions on the classical phase space $T^*Q$, associated with a given spacetime $Q$, is a commutative algebra. According to Dirac, quantization can be achieved by replacing a function $f$ in this algebra by an operator $ \hat f $ and equating $i\hbar$ times the Poisson bracket between functions to the commutator between the corresponding operators. In classical physics, the functions $f$ commute, so $ {\mathcal{F}}(T^*Q)$ is a commutative algebra. But the corresponding quantum algebra $\hat{\mathcal{F}}$ is not commutative. Dynamics is on $\hat{\mathcal{F}}$. So quantum physics is $\textit{noncommutative dynamics}$.

A particular aspect of this dynamics is $\textit{fuzzy}$ $\textit{phase space}$ where we cannot localize points, and which has an attendent effective ultraviolet cutoff. A fuzzy phase space can still admit the action of a continuous symmetry group such as the spatial rotation group as the automorphism group \cite{madore}. For example, one can quantize functions on a sphere $S^{2}$ to obtain a fuzzy sphere \cite{fuzzybook}. It admits $SO(3)$ as an automorphism group. The fuzzy sphere can be identified with the algebra $M_{n}$ of $n \times n$ complex matrices. The volume of phase space in this case becomes finite. Semiclassically there are a finite number of cells on the fuzzy sphere, each with a finite area \cite{madore}.
  
Thus in quantum physics, the commutative algebra of functions on phase space is deformed to a noncommutative algebra, leading to a ``noncommutative phase space''. Such deformations, characteristic of quantization, are now appearing in different approaches to fundamental physics. Examples are the following: 

1.) Noncommutative geometry has made its appearance as a method for regularizing quantum field theories (qft's) and in studies of deformation quantization.

2.) It has turned up in string physics as quantized $D$-branes.

3.) Certain approaches to canonical gravity \cite{paulo} have used noncommutative geometry with great effectiveness.

4.) There are also plausible arguments based on the uncertainty principle \cite{doplicher} that indicate a noncommutative spacetime in the presence of gravity.

5.) It has been conjuctered by `t Hooft \cite{thooft} that the horizon of a black hole should have a fuzzy 2-sphere structure to give a finite entropy.

6.) A noncommutative structure emerges naturally in quantum Hall effect \cite{qhe}.

\section{Noncommutative Spacetime}
\subsection{A Little Bit of History}
The idea that spacetime geometry may be noncommutative is old. It goes back to Schr\"odinger and to Heisenberg who raises this possibilty in a letter to Rudolph Peierls in the 30's. Heisenberg complained in this letter that he did not know enough mathematics to explore the physical consequences of this possibilty. Peierls mentioned Heisenberg's ideas to Wolfgang Pauli. Pauli in turn explained it to Hartland Snyder. In 1947 Snyder used the noncommutative structure of spacetime to introduce a small length scale cut-off in field theory without breaking Lorentz invariance \cite{Snyder}. In the same year, Yang \cite{yang} also published a paper on quantized spacetime, extending Snyder's work. The term `noncommutative geometry' was introduced by von Neumann \cite{madore}. He used it to describe in general a geometry in which the algebra of noncommuting linear operators replaces the algebra of functions. 

Snyder's idea was forgotten with the successful development of the renormalization program. Later, in the 1980's Connes \cite{connesA} and Woronowicz \cite{woronowicz} revived noncommutative geometry by introducing a differential structure in the noncommutative framework.

We should also mention the role of Joe Weinberg in these developments. Joe was  a student of Robert Oppenheimer and was a close associate of Wolfgang Pauli and a classmate of Julian Schwinger. He was the person accused of passing nuclear secrets to the Soviets and who lost his job in 1952 at the University of Minnesota for that reason. His wife supported the family for several years. Eventually he got a faculty position at Case Western Reserve University in 1958 and from there, he came to Syracuse University.

Joe was remarkable. He seemed to know everything, from Sanskrit to noncommutative geometry, and published very little. He had done extensive research on this new vision of spacetime. His manuscripts are preserved in the Syracuse University archives.
 
\subsection{Spacetime Uncertaintities}
It is generally believed that the picture of spacetime as a manifold of points breaks down at distance scales of the order of the Planck length: Spacetime events cannot be localized with an accuracy given by Planck length.  

The following argument can be found in Doplicher {\it et al.} \cite{doplicher}. In order to probe physics at a fundamental length scale $L$ close to the Planck scale, the Compton wavelength $\frac{\hbar}{Mc}$ of the probe must fulfill
\begin{equation}
\frac{\hbar}{Mc}\,\leq\, L\ \ \textrm{or}\ \
M\,\geq\,\frac{\hbar}{Lc}\,\simeq\,\textrm{Planck mass}.
\end{equation}
Such high mass in the small volume $L^3$ will strongly affect gravity and can cause black holes and their horizons to form. This suggests a fundamental length limiting spatial localization. That is, there is a space-space uncertainty,
\beq
\Delta x_{1}\Delta x_{2} + \Delta x_{2}\Delta x_{3} + \Delta x_{3}\Delta x_{1} \ \gtrsim L^{2}
\eeq

Similar arguments can be made about time localization. Observation of very short time scales requires very high energies. They can produce black holes and black hole horizons will then limit spatial resolution suggesting 
\beq
\Delta x_{0}(\Delta x_{1} + \Delta x_{2} + \Delta x_{3}) \geq L^{2}.
\eeq

The above uncertainty relations suggest that spacetime ought to be described as a noncommutative manifold just as classical phase space is replaced by noncommutative phase space in quantum physics which leads to Heisenberg's uncertainty relations. The points on the classical commutative manifold should then be replaced by states on a noncommutative algebra. 
\subsection{The Groenewold-Moyal Plane}
The noncommutative Groenewold-Moyal (GM) spacetime is a deformation of ordinary spacetime in which the spacetime coordinate functions $\widehat{x}_\mu$ do not commute \cite{ConnesC, Madore, Landi, Bondia}: 
\beq
\label{non-commu-coordinates}
[\widehat{x}^{\mu}, \widehat{x}^{\nu}]=i \theta^{\mu \nu},~~~\theta^{\mu \nu}=-\theta^{\nu \mu}=\textrm{constants},
\eeq
where the coordinate functions $\widehat{x}_\mu$ give Cartesian coordinates $x_{\mu}$ of (flat) spacetime:
\beq
\widehat{x}_{\mu}(x)=x_{\mu}.
\eeq
The deformation matrix $\theta$ is taken to be a real and antisymmetric constant matrix \cite{RGCai}. Its elements have the dimension of (length)$^{2}$, thus a scale for the smallest patch of area in the ${\mu}$ - ${\nu}$ plane. They also give a measure of the strength of noncommutativity. One cannot probe spacetime with a resolution below this scale. That is, spacetime is ``fuzzy'' \cite{Ydri} below this scale. In the limit $\theta_{\mu \nu} \to 0$, one recovers ordinary spacetime.
\section{The Star Products}
In this part we will go into more details of the GM plane. The GM plane incorporates spacetime uncertainties. Such an introduction of spacetime noncommutativity replaces point-by-point multiplication of two fields by a type of ``smeared'' product. This type of product is called a star product.  
\subsection{Deforming an Algebra}
There is a general way of deforming the algebra of functions on a manifold $M$ \cite{queiroz}. The GM plane, ${\cal A}_{\theta}({\mathbb R}^{d+1})$, associated with spacetime ${\mathbb R}^{d+1}$ is an example of such a deformed algebra. 

Consider a Riemannian manifold $(M,g)$ with metric $g$. If the group $\R^N\;(N\geq 2)$ acts as a group of isometries on $M$, then it acts on the Hilbert space $L^2(M,d\mu_g)$ of square integrable functions on $M$. The volume form $d\mu_g$ for the scalar product on $L^2(M,d\mu_g)$ is induced from $g$.

If $\Big\{\lambda = (\lambda_1, \ldots, \lambda_N)\Big\}$ denote the unitary irreducible representations (UIR's) of $\R^N$, then we can write
\begin{equation}\label{eq:1}
L^2(M, d\mu_g) = \bigoplus_\lambda {\cal H}^{(\lambda)} \;,
\end{equation}
where $\R^N$ acts by the UIR $\lambda$ on ${\cal H}^{(\lambda)}$. 

We choose $\lambda$ such that
\beq
\lambda : a \longrightarrow e^{i \lambda a}
\eeq
where $a=(a_{1}, a_{2}, \cdots, a_{N}) \in {\mathbb R}^{N}$. 

Choose two smooth functions $f_{\lambda}$ and $f_{\lambda '}$ in ${\cal H}^{(\lambda)}$ and ${\cal H}^{(\lambda ')}$. Then under the pointwise multiplication
\beq
f_{\lambda} \otimes f_{\lambda '} \rightarrow f_{\lambda} f_{\lambda '}
\eeq
where, if $p$ is a point on $M$,
\beq
(f_{\lambda} f_{\lambda '})(p) = f_{\lambda}(p) f_{\lambda '}(p).
\eeq
Also
\beq
\label{eq:2}
f_{\lambda} f_{\lambda '} \in {\cal H}^{(\lambda + \lambda ')}
\eeq
where we have taken the group law as addition.

Let $\theta^{\mu \nu}$ be an antisymmetric constant matrix in the space of UIR's of ${\mathbb R}^N$. The above algebra with pointwise multiplication can be deformed into a new deformed algebra. The pointwise product becomes a $\theta$ dependent ``smeared" product $*_{\theta}$ in the deformed algebra,
\beq
\label{eq:3}
f_\lambda *_\theta f_{\lambda'} =  f_\lambda \;  f_{\lambda'} \; e^{-\frac{i}{2} \lambda_\mu \theta^{\mu \nu} \lambda'_\nu}\;.
\eeq
This deformed algebra is also associative because of eqn. (\ref{eq:2}). The GM plane, ${\cal A}_\theta({\mathbb R}^{d+1})$, is a special case of this algebra.

In the case of the GM plane, the group ${\mathbb R}^{d+1}$ acts on ${\cal A}_{\theta}({\mathbb R}^{d+1})$ $\{={\cal C}^\infty({\mathbb R}^{d+1}) \; \textrm{as a set} \}$ by translations leaving the flat Euclidean metric invariant. The IRR's are labelled by the ``momenta" $\lambda = p = (p^{0}, p^{1}, \ldots, p^{d})$. A basis for the Hilbert space ${\cal H}^{(p)}$ is formed by plane waves $e_{p}$ with $e_{p}(x) = e^{-ip_{\mu}  x^{\mu}}$, $x = (x^0,x^1, \ldots, x^d)$ being a point of ${\mathbb R}^{d+1}$. The $*$-product for the GM plane follows from eqn. (\ref{eq:3}),
\beq
\label{eq:4}
e_{p} \ast_\theta e_{q} = e_{p} \; e_{q} \; e^{-\frac{i}{2}p_\mu \theta^{\mu \nu}q_\nu}\;.
\eeq
This $*$-product defines the Moyal plane ${\cal A}_\theta({\mathbb R}^{d+1})$.

In the limit $\theta_{\mu \nu} \rightarrow 0$, the operators $e_{p}$  and $e_{q}$ become commutative functions on ${\mathbb R}^{N}$.

\subsection{The Voros and Moyal Star Products}
This section is based on the book \cite{fuzzybook}.

The algebra ${\cal A}_{0}$ of smooth functions on a manifold $M$ under point-wise multiplication is a  commutative algebra. In the previous section we saw that ${\cal A}_{0}$ can be deformed into a new algebra ${\cal A}_{\theta}$ in which the point-wise product is deformed to a noncommutative (but still associative) product called the $*$-product. 

Such deformations were studied by Weyl, Wigner, Groenewold and Moyal \cite{lipkin, weyl, grosse-pres}. The $*$-product has a central role in many discussions of noncommutative geometry. It appears in other branches of physics like quantum optics.

The $*$-product can be obtained from the algebra of creation and annihilation operators. It is explained below.
\subsubsection{Coherent States}
The dynamics of a quantum harmonic oscillator most closely resembles that of a classical harmonic oscillator when the oscillator quantum state is a coherent state. Consider a quantum oscillator with annihilation and creation operators $a$, $a^\dagger$. The coherent states are
\beq
|z \rangle = e^{z a^\dagger - {\bar z} a} |0 \rangle = e^{-\frac{1}{2}|z|^2} e^{z a^\dagger} |0 \rangle \,, \quad z \in {\mathbb C}.\nn
\eeq
They have the properties
\beq
a |z \rangle = z |z \rangle \,; \quad \quad \langle z^\prime | z \rangle = e^{\frac{1}{2} |z - z^\prime|^2} \,.
\label{eq:csp1}
\eeq

The coherent states are overcomplete, with the resolution of identity
\beq
{\bf 1} = \int \frac{d^2 z}{\pi} |z \rangle \langle z|  \,, \quad d^2 z
= dx_1 dx_2 \,,
\label{eq:overcomp}
\eeq
where
\beq
z=\frac{x_1 + i x_2}{\sqrt2} \,.\nn
\eeq

Consider an operator ${\hat A}$. The ``symbol" of ${\hat A}$ is a function $A$ on ${\mathbb C}$ with values $A ( z \,, {\bar z}) = \langle z| {\hat A} | z \rangle$. A central property of coherent states is that an operator ${\hat A}$ is determined just by its diagonal matrix elements, that is, by the symbol $A$ of ${\hat A}$.
\subsubsection{The Coherent State or Voros $*$-product on the GM Plane}
As indicated above, we can map an operator $\hat A$ to a function $A$ using coherent states as follows:
\beq
{\hat A} \longrightarrow A \,, \quad A(z \,, {\bar z}) = \langle z | {\hat
  A} |z \rangle.
\eeq
This is a bijective linear map and induces a product $*_C$ on functions ($C$ indicating ``coherent state''). With this product, we get an algebra $(C^\infty({\mathbb C}) \,, *_C)$ of functions. Since the map ${\hat A} \rightarrow A$ has the property $({\hat A})^* \rightarrow A^* \equiv {\bar A}$, this map is a $*$-morphism from operators to $(C^\infty({\mathbb C}) \,, *_C)$ where $*$ on functions is complex conjugation.

Let us get familiar with this new function algebra. 

The image of $a$ is the function $\alpha$ where $\alpha(z\,,{\bar z}) =z$. The image of $a^n$ has the value $z^n$ at $(z \,, {\bar z})$, so by definition,
\beq
(\alpha *_C \alpha \ldots *_C \alpha) (z \,, {\bar z}) = z^n \,.
\eeq

The image of $a^* \equiv a^\dagger$ is ${\bar \alpha}$ where ${\bar \alpha}(z, {\bar z}) = {\bar z}$ and that of $(a^*)^n$ is ${\bar \alpha} *_C {\bar \alpha} \cdots *_C {\bar \alpha}$ where
\beq
{\bar \alpha} *_C {\bar \alpha} \cdots *_C {\bar \alpha}(z\,, {\bar
  z}) = {\bar z}^n \,.
\eeq

Since $\langle z | a^* a | z \rangle = {\bar z} z$ and $\langle z | a a^* | z \rangle = {\bar z} z + 1$, we get
\beq
{\bar \alpha} *_C \alpha = {\bar \alpha} \alpha \,, \quad \quad
\alpha *_C {\bar \alpha} = \alpha {\bar \alpha} + {\bf 1} \,,
\eeq
where $ {\bar \alpha} \alpha =  \alpha {\bar \alpha}$ is the pointwise product of $\alpha$ and ${\bar \alpha}$, and ${\bf 1}$ is the constant function with value $1$ for all $z$. 

For general operators ${\hat f}$, the construction proceeds as follows. Consider
\beq
: e^{\xi a^\dagger - {\bar \xi} a}:
\eeq
where the normal ordering symbol $: \cdots :$ means as usual that $a^\dagger$'s are to be put to the left of $a$'s. Thus
\bea
: a a^\dagger a^\dagger a : &=& a^\dagger a^\dagger a a \,, \nonumber \\
: e^{\xi a^\dagger - {\bar \xi} a}: &=& e^{\xi a^\dagger} e^{-{\bar
      \xi} a} \,. \nn
\eea

Hence
\beq
\langle z | :e^{\xi a^\dagger - {\bar \xi} a}: |z \rangle = e^{\xi
  {\bar z} - {\bar \xi} z} \,.
\eeq
 
Writing ${\hat f}$ as a Fourier transform,
\beq
{\hat f} = \int \frac{d^2 \xi}{\pi} : e^{\xi a^\dagger - {\bar \xi}
  a}: {\tilde f}(\xi \,, {\bar \xi}) \,, \quad \quad {\tilde f} 
(\xi \,, {\bar \xi}) \in {\mathbb C} \,,
\eeq
its symbol is seen to be
\beq
f = \int \frac{d^2 \xi}{\pi} e^{\xi {\bar z} - {\bar \xi} z} {\tilde
  f}(\xi \,, {\bar \xi}) \,.
\eeq
    
This map is invertible since $f$ determines ${\tilde f}$. Consider also the second operator
\beq
{\hat g} = \int \frac{d^2 \eta}{\pi} : e^{\eta a^\dagger - {\bar \eta}
  a}: {\tilde g}(\eta \,, \bar {\eta}) \,,
\eeq
and its symbol
\beq
g = \int \frac{d^2 \eta}{\pi} e^{\eta {\bar z} - {\bar \eta} z}
{\tilde g}(\eta \,, \bar {\eta}) \,. 
\eeq
 
The task is to find the symbol $f *_C g$ of ${\hat f}{\hat g}$. Let us first find
\beq
e^{\xi {\bar z} - {\bar \xi} z} *_C  e^{\eta {\bar z} - {\bar \eta} z} \,.
\eeq

We have
\beq
:e^{\xi a^\dagger - {\bar \xi} a}: \,  : e^{\eta a^\dagger - {\bar
      \eta} a}: = : e^{\xi a^\dagger - {\bar \xi} a} \,  
e^{\eta a^\dagger - {\bar \eta} a}: e^{-{\bar \xi}{\eta}}
\eeq
and hence
\bea
\label{eq:exp1}
e^{\xi {\bar z} - {\bar \xi} z} *_C e^{\eta {\bar z} - {\bar \eta} z}
&=& e^{-{\bar \xi} \eta} e^{\xi {\bar z} - {\bar \xi} z} \,  
e^{\eta {\bar z} - {\bar \eta} z} \nonumber \\
&=& e^{\xi {\bar z} - {\bar \xi} z} e^{{\overleftarrow \partial}_z \,
  {\overrightarrow \partial}_{\bar z}} 
e^{\eta {\bar z} - {\bar \eta} z} \,.
\eea
 
The bidifferential operators $\big ({\overleftarrow \partial}_z \, {\overrightarrow \partial}_{\bar z} \big )^k \,, (k= 1,2,...)$ have the definition
\beq
\alpha \big ({\overleftarrow \partial}_z \, {\overrightarrow
  \partial}_{\bar z} \big )^k \beta \, (z \,, {\bar z}) =  
\frac{\partial^k \alpha (z \,, {\bar z})}{\partial z^k}
\frac{\partial^k \beta (z \,, {\bar z})}{\partial {\bar z}^k} \,.
\eeq

The exponential in (\ref{eq:exp1}) involving them can be defined using the power series.

The coherent state $*$-product $f *_C g$ follows from (\ref{eq:exp1}):
\beq
f *_C g \,(z \,, {\bar z}) = \big ( f  e^{{\overleftarrow \partial}_z
  \, {\overrightarrow \partial}_{\bar z}} g \big ) (z \,, {\bar z})  \,.    
\label{eq:csstar1}
\eeq

We can explicitly introduce a deformation parameter $\theta > 0 $ in the discussion by changing (\ref{eq:csstar1}) to
\beq
f *_C g \, (z \,, {\bar z}) = \big ( f  e^{ \theta \, {\overleftarrow
    \partial}_z \, {\overrightarrow \partial}_{\bar z}} g \big )  
(z \,, {\bar z}) \,.   
\label{eq:csstar2}
\eeq
  
After rescaling $z^\prime = \frac{z}{\sqrt{\theta}}$, (\ref{eq:csstar2}) gives (\ref{eq:csstar1}). As $z^\prime$ and ${\bar z}^\prime$ after quantization become $a \,, a^\dagger$, $z$ and ${\bar z}$ become the scaled oscillators $a_\theta \,, a_\theta^\dagger$
\beq
\lbrack a_\theta \,, a_\theta \rbrack = \lbrack a_\theta^\dagger  \,,
a_\theta^\dagger \rbrack = 0 \,, \quad  \lbrack a_\theta \,,
a_\theta^\dagger \rbrack = \theta \,. 
\label{eq:tetacom}
\eeq

Equation (\ref{eq:tetacom}) is associated with the Moyal plane with Cartesian coordinate functions $x_1 \,, x_2$. If $a_\theta = \frac{x_1 + i x_2}{\sqrt2} \,, a_\theta^\dagger = \frac{x_1 - i x_2}{\sqrt2}$,
\beq
\lbrack x_i \,, x_j \rbrack = i \theta \varepsilon_{ij} \,, \quad
\varepsilon_{ij} = - \varepsilon_{ji} \,, \quad \varepsilon_{12} = 1 \,.
\label{eq:deform1}
\eeq

The Moyal plane is the plane ${\mathbb R}^2$, but with its function algebra deformed in accordance with eqn. (\ref{eq:deform1}). The deformed algebra has the product eqn. (\ref{eq:csstar2}) or equivalently the Moyal product derived below.
\subsubsection{The Moyal Product on the GM Plane}
We get this by changing the map ${\hat f} \rightarrow f$ from operators to functions. For a given function $f$, the operator ${\hat f}$ is thus different for the coherent state and Moyal $*$'s. The $*$-product on two functions is accordingly also different.

Let us introduce the Weyl map and the Weyl symbol. The Weyl map of the operator
\beq 
{\hat f} = \int \frac{d^2 \xi}{\pi} {\tilde f}(\xi \,, {\bar \xi})
e^{\xi a^\dagger - {\bar \xi} a}
\label{eq:Weyl1}
\eeq
to the function $f$ is defined by
\beq
f(z\,,{\bar z}) = \int \frac{d^2 \xi}{\pi} {\tilde f}(\xi \,, 
{\bar \xi}) e^{\xi {\bar z} - {\bar \xi} z} \,. 
\label{eq:Weyl2}
\eeq

Equation (\ref{eq:Weyl2}) makes sense since ${\tilde f}$ is fully determined by ${\hat f}$ as follows:
\beq
\langle z| {\hat f} | z \rangle = \int \frac{d^2 \xi}{\pi} {\tilde
  f}(\xi \,, \bar {\xi}) e^{-\frac{1}{2} \xi {\bar \xi} } 
e^{\xi {\bar z} - {\bar \xi} z} \,.    \nn   
\eeq
${\tilde f}$ can be calculated from here by Fourier transformation.

The map is invertible since ${\tilde f}$ follows from $f$ by the Fourier transform of eqn. (\ref{eq:Weyl2}) and ${\tilde f}$ fixes ${\hat f}$ by eqn. (\ref{eq:Weyl1}). $f$ is called the {\it Weyl symbol} of ${\hat f}$.

As the Weyl map is bijective, we can find a new $*$ product, call it $*_W$, between functions by setting $f*_W g = \, \mbox{Weyl symbol of} \, {\hat f}{\hat g}$.

For
\beq
{\hat f}(\xi, \bar{\xi}) =  e^{\xi a^\dagger - {\bar \xi} a} \,, \quad {\hat g}(\eta, \bar{\eta}) =
e^{\eta a^\dagger - {\bar \eta} a} \,,\nn 
\eeq
to find $f*_W g$, we first rewrite ${\hat f}{\hat g}$ according to
\beq
{\hat f}{\hat g} = e^{\frac{1}{2}(\xi {\bar \eta} - {\bar \xi} \eta)}  e^{(\xi +
  \eta) a^\dagger - ({\bar \xi} +{\bar \eta}) a} \,.\nn  
\eeq
Hence
\bea
f*_W g \,(z\,, {\bar z}) &=& e^{\xi {\bar z}-{\bar \xi} z}  
e^{\frac{1}{2}(\xi {\bar \eta} - {\bar \xi} \eta)} 
e^{\eta {\bar z}-{\bar \eta} z}
\nonumber \\     
&=& f e^{\frac{1}{2} \big ( {\overleftarrow \partial}_z \,
{\overrightarrow \partial}_{\bar z} - {\overleftarrow \partial}_{\bar
  z} \, {\overrightarrow \partial}_z
\big )} g \, (z \,,{\bar z}) \,. 
\label{eq:Weyl3}
\eea
 
Multiplying by ${\tilde f}$, ${\tilde g}$ and integrating, we get eqn. (\ref{eq:Weyl3}) for arbitrary functions:
\beq
f*_W g \, (z\,, {\bar z}) = \Big ( f e^{\frac{1}{2} \big (
  {\overleftarrow \partial}_z \, {\overrightarrow \partial}_{\bar z} - 
{\overleftarrow \partial}_{\bar z} \, {\overrightarrow \partial}_z
\big )} g \Big ) (z \,,{\bar z}) \,.
\eeq
 
Note that
\beq
{\overleftarrow \partial}_z \, {\overrightarrow \partial}_{\bar z}
-{\overleftarrow \partial}_{\bar z} \, {\overrightarrow \partial}_z 
= i ( {\overleftarrow \partial}_1 \, {\overrightarrow \partial}_2
-{\overleftarrow \partial}_2 \, {\overrightarrow \partial}_1 ) 
= i \varepsilon_{ij}  {\overleftarrow \partial}_i \, {\overrightarrow
  \partial}_j  \,.\nn 
\eeq

Introducing also $\theta$, we can write the $*_W$-product as
\beq
f *_W g = f e^{i \frac{\theta}{2} \varepsilon_{ij}  {\overleftarrow \partial}_i
  \, {\overrightarrow \partial}_j} g \,. 
\label{eq:Weyl4}
\eeq
 
By eqn. (\ref{eq:deform1}), $\theta \varepsilon_{ij} = \omega_{ij}$ fixes the Poisson brackets, or the Poisson structure on the Moyal plane. Eqn. (\ref{eq:Weyl4}) is customarily written as
\beq
f *_W g = f  e^{\frac{i}{2} \omega_{ij}  {\overleftarrow \partial}_i \,
  {\overrightarrow \partial}_j} g \nn
\eeq
using the Poisson structure. (But we have not cared to position the indices so as to indicate their tensor nature and to write $\omega^{ij}$.)
\subsection{Properties of $*$-Products}
A $*$-product without a subscript indicates that it can be either a $*_C$ or a $*_W$.
\subsubsection{Cyclic Invariance}
The trace of operators has the fundamental property $Tr {\hat A} {\hat B} = Tr {\hat B} {\hat A}$, which leads to the general cyclic identities
\beq
Tr \, {\hat A}_1 \ldots {\hat A}_n = Tr \, {\hat A}_n {\hat A}_1
\ldots {\hat A}_{n-1} \,. 
\label{eq:ctr1}
\eeq
 
We now show that
\beq
Tr \, {\hat A} {\hat B} = \int \frac{d^2 z}{\pi} \,  A * B \, (z \,,
{\bar z}) \,, \quad \quad * = *_C \quad \mbox{or} \quad *_W \,. 
\label{eq:ctr2}
\eeq
   
(The functions on the right hand side are different for $*_C$ and $*_W$ if ${\hat A} \,, {\hat B}$ are fixed). From this follows the analogue of (\ref{eq:ctr1}):
\beq
\int \frac{d^2 z}{\pi} \, \big ( A_1 * A_2 * \cdots * A_n) \, (z \,, {\bar
  z} \big ) = \int \frac{d^2 z}{\pi} \big ( A_n * A_1 * \cdots * 
A_{n-1}) \, (z \,, {\bar z} \big ) \,. 
\label{eq:ctr3}
\eeq

For $*_C$, eqn. (\ref{eq:ctr2}) follows from eqn. (\ref{eq:overcomp}). The coherent state image of $ e^{\xi a^\dagger - {\bar \xi} a} $ is the function with value
\beq
e^{\xi {\bar z} - {\bar \xi} z} e^{-\frac{1}{2}{\bar \xi}{\xi}}
\label{eq:csf1}
\eeq 
at $z$, with a similar correspondence if $\xi \rightarrow \eta$. So
\beq
Tr  \, e^{\xi a^\dagger - {\bar \xi} a} \, e^{\eta a^\dagger - {\bar
    \eta} a} = \int {\frac{d^2 z}{\pi}} \, \Big ( 
e^{\xi {\bar z} - {\bar \xi} z} e^{-\frac{1}{2}{\bar \xi}{\xi}} \Big)
    \Big ( e^{\eta {\bar z} - {\bar \eta} z}  
e^{-\frac{1}{2}{\bar \eta}{\eta}} \Big ) e^{-{\bar \xi}{\eta}}\nn
\eeq

The integral produces the $\delta$-function
\beq
\prod_i 2 \delta (\xi_i + \eta_i) \,, \quad \quad  \xi_i = \frac{\xi_1 +
  \xi_2}{\sqrt{2}} \,, \quad \eta_i = \frac{\eta_1 + \eta_2} {\sqrt{2}} \,.\nn
\eeq

We can hence substitute $e^{- \big ( \frac{1}{2}{\bar \xi}{\xi} +  \frac{1}{2}{\bar \eta} {\eta} + {\bar \xi}{\eta} \big)}$ by $e^{\frac{1}{2} (\xi {\bar \eta} - {\bar \xi} \eta)}$ and get eqn. (\ref{eq:ctr2}) for Weyl $*$ for these exponentials and so for general functions by using  eqn. (\ref{eq:Weyl1}).
\subsubsection{A Special Identity for the Weyl Star}
The above calculation also gives the identity
\beq
\int \frac{d^2 z}{\pi} A *_W B \, (z \,, {\bar z}) = \int \frac{d^2
  z}{\pi} A (z \,, {\bar z}) \, B \, (z \,, {\bar z}) \,.\nn  
\eeq

That is because
\beq
\prod_i \delta(\xi_i + \eta_i) \, e^{\frac{1}{2} (\xi {\bar \eta} -
{\bar \xi} \eta)} = \prod_i \, \delta(\xi_i + \eta_i) \,.\nn  
\eeq

In eqn. (\ref{eq:ctr3}), $A$ and $B$ in turn can be Weyl $*$-products of other functions. Thus in integrals of Weyl $*$-products of functions, one $*_W$ can be replaced by the pointwise (commutative) product:
\bea
&&\int \frac{d^2 z}{\pi} \big ( A_1 *_W A_2 *_W \cdots A_K \big ) *_W
  ( B_1 *_W B_2 *_W \cdots B_L \big ) \, (z \,, {\bar z})  
\nonumber \\
&& \quad \quad \quad \quad = \int \frac{d^2 z}{\pi} \big ( A_1 *_W A_2
*_W \cdots A_K \big ) \,( B_1 *_W B_2 *_W  \cdots B_L \big ) \, (z \,,
{\bar z}) \,.\nn  
\eea

This identity is frequently useful.
\subsubsection{Equivalence of $*_C$ and $*_W$}
For the operator
\beq
{\hat A} = e^{\xi a^\dagger -{\bar \xi} a} \,,
\eeq
the coherent state function $A_C$ has the value (\ref{eq:csf1}) at $z$, and the Weyl symbol $A_W$ has the value
\beq
A_W(z \,, {\bar z}) = e^{\xi {\bar z} - {\bar \xi} z} \,.\nn
\eeq

As both $\big ( C^\infty({\mathbb R}^2) \,, *_C \big )$ and $\big (C^\infty({\mathbb R}^2) \,, *_W \big )$ are isomorphic to the operator algebra, they too are isomorphic. The isomorphism is established by the maps
\beq
A_C \longleftrightarrow A_W \nn
\eeq
and their extension via Fourier transform to all operators and functions ${\hat A} \,, A_{C \,, W}$.

Clearly
\beq
A_W = e^{-\frac{1}{2} \partial_z \partial_{\bar z}} A_C \,, \quad 
A_C = e^{\frac{1}{2} \partial_z \partial_{\bar z}} A_W  \,, \nonumber \\
A_C *_C B_C \longleftrightarrow A_W *_W B_W \,.\nn  
\eeq

The mutual isomorphism of these three algebras is a $*$-isomorphism since 
$({\hat A} {\hat B})^\dagger \longrightarrow {\bar B}_{C \,, W} *_{C \,, W} {\bar A}_{C \,, W}$. 
\subsubsection{Integration and Tracial States}
This is a good point to introduce the ideas of a state and a tracial state on a $*$-algebra ${\cal A}$ with unity ${\bf 1}$.

A state $\omega$ is a linear map from ${\cal A}$ to ${\mathbb C}$, $\omega (a) \in {\mathbb C}$ for all $a \in {\cal A}$ with the following properties:
\bea
\omega(a^*) &=& \overline{\omega(a)} \,, \nonumber \\  
\omega (a^*a) & \geq & 0 \,, \nonumber \\
\omega({\bf 1}) & = & 1 \,.\nn
\label{eq:ts1}
\eea

If ${\cal A}$ consists of operators on a Hilbert space and $\rho$ is a density matrix, it defines a state $\omega_\rho$ via
\beq
\omega_\rho (a) =  Tr (\rho a) \,.
\label{eq:ts2}
\eeq

If $\rho = e^{- \beta H}/ Tr (e^{-\beta H})$ for a Hamiltonian $H$, it gives a Gibbs state via eqn. (\ref{eq:ts2}).

Thus the concept of a state on an algebra ${\cal A}$ generalizes the notion of a density matrix. There is a remarkable construction, the Gel'fand- Naimark-Segal (GNS) construction, which shows how to associate any state with a rank-$1$ density matrix \cite{Haag}.

A state is {\it tracial} if it has cyclic invariance:
\beq
\omega (ab) = \omega (ba) \,.
\label{eq:tracial1}
\eeq

The Gibbs state is not tracial, but fulfills an identity generalizing eqn. (\ref{eq:tracial1}). It is a Kubo-Martin-Schwinger (KMS) state \cite{Haag}. 

A positive map $\omega^\prime$ is in general an unnormalized state: It must fulfill all the conditions that a state fulfills, but is not obliged to fulfill the condition $\omega^\prime({\bf 1}) = 1$.

Let us define a positive map $\omega^\prime$ on $(C^\infty({\mathbb R}^{2}) \,, *)$ ($*
= *_C \, \mbox{or} \, *_W$) using integration:
\beq
\omega^\prime(A) = \int \frac{d^2 z}{\pi} \, {\hat A}
(z \,, {\bar z}) \,.\nn
\eeq

It is easy to verfy that $\omega^\prime$ fulfills the properties of a positive map. A {\it tracial} positive map $\omega^\prime$ also has the cyclic invariance, eqn. (\ref{eq:tracial1}).

The cyclic invariance (\ref{eq:tracial1}) of $\omega^\prime (A * B)$ means that
it is a tracial positive map.

\subsubsection{The $\theta$-Expansion}

On introducing $\theta$, we have (\ref{eq:csstar2}) and
\beq
f *_W g (z \,, {\bar z}) =  f e^{\frac{\theta}{2} \big (
  {\overleftarrow \partial}_z \, {\overrightarrow \partial}_{\bar z} - 
{\overleftarrow \partial}_{\bar z} \, {\overrightarrow \partial}_z
  \big )} g \, (z \,,{\bar z}) \,.\nn 
\eeq
   
The series expansion in $\theta$ is thus
\beq
f *_C g \, (z \,, {\bar z}) = f g \, (z \,, {\bar z}) + \theta \,
\frac{\partial f}{\partial z} (z \,, {\bar z}) \frac{\partial g} 
{\partial {\bar z}} (z \,, {\bar z}) + {\cal O} (\theta^2) \,,\nn
\eeq
\beq
f *_W g \, (z \,, {\bar z}) =  f g (z \,, {\bar z}) + \frac{\theta}{2}
\Big ( \frac{\partial f}{\partial z} \frac{\partial g} 
{\partial {\bar z}} - \frac{\partial f}{\partial {\bar z}}
\frac{\partial g} {\partial z} \Big ) \, (z \,, {\bar z}) +  {\cal O} (\theta^2) \,.\nn
\eeq

Introducing the notation
\beq
\lbrack f \,, g \rbrack_* = f * g - g * f \,, \quad *=*C \quad
\mbox{or} \quad *_W \,, 
\label{eq:ps1}
\eeq
we see that
\bea
\lbrack f \,, g \rbrack_{*_C} &=& \theta \Big ( \frac{\partial
  f}{\partial z} \frac{\partial g} {\partial {\bar z}} -  
\frac{\partial f}{\partial {\bar z}} \frac{\partial g} {\partial z}
\Big ) (z \,, {\bar z}) +  {\cal O} (\theta^2) \,, \nonumber \\ 
\lbrack f \,, g \rbrack_{*_W} &=& \theta \Big ( \frac{\partial
  f}{\partial z} \frac{\partial g} {\partial {\bar z}} -  
\frac{\partial f}{\partial {\bar z}} \frac{\partial g} {\partial z}
\Big ) (z \,, {\bar z}) +  {\cal O} (\theta^2) \,.\nn       
\eea
 
We thus see that
\beq
\lbrack f \,, g \rbrack_* = i \theta \{f \,, g \}_{P.B.} + {\cal O} (\theta^2) \,,
\label{eq:ps2}
\eeq
where $\{f \,, g \}$ is the Poisson bracket of $f$ and $g$ and the ${\cal O}(\theta^2)$ term depends on $*_{C \,, W}$. Thus the $*$-product is an associative product which to leading order in the deformation parameter (``Planck's constant") $\theta$ is compatible with the rules of quantization of Dirac. We can say that with the $*$-product, we have deformation quantization of the classical commutative algebra of functions.

But it should be emphasized that even to leading order in $\theta$, $f*_C g$ and $f *_W g$ do not agree. Still the algebras $\big (C^\infty({\mathbb R}^2 \,, *_C) \big )$ and $\big ( C^\infty({\mathbb R}^2 \,, *_W) \big )$ are $*$-isomorphic.

If a Poisson structure on a manifold $M$ with Poisson bracket $\{. \,, .\}$ is given, then one can have a $*$-product $f * g$ as a formal power series in $\theta$ such that eqn. (\ref{eq:ps2}) holds \cite{Kontsevich}.
\section{Spacetime Symmetries on Noncommutative Plane}
In this section we address how to implement spacetime symmetries on the noncommutative spacetime algebra ${\cal A}_{\theta}({\mathbb R}^{N})$, where functions are multiplied by a $*$-product. In section 2, we modelled the spacetime noncommutativity using the commutation relations given by eqn. (\ref{non-commu-coordinates}). Those relations are clearly not invariant under naive Lorentz transformations. That is, the noncommutative structure we have modelled breaks Lorentz symmetry. Fortunately, there is a way to overcome this difficulty: one can interpret these relations in a Lorentz-invariant way by implementing a deformed Lorentz group action \cite{chaichian}.
\subsection{The Deformed Poincar\'e Group Action}
The single particle states in quantum mechanics can be identified with the carrier space of the one-particle unitary irreducible representations (UIRR's) of the identity component of the Poincar\'e group, $P_{+}^{\uparrow}$ or rather its two-fold cover $\bar{P}^{\uparrow}_{+}$. Let $U(g)$, $g \in \bar{P}^{\uparrow}_{+}$, be the UIRR for a spinless particle of mass $m$ on a Hilbert space ${\cal H}$. Then ${\cal H}$ has the basis $\{ |k \rangle\}$ of momentum eigenstates, where $k=(k_{0}, {\bf k})$, $k_{0}=|\sqrt{{\bf k}^{2}+m^{2}}|$. $U(g)$ transforms $|k\rangle$ according to 
\beq
U(g) |k\rangle = |g k\rangle.
\eeq
Then conventionally $\bar{P}^{\uparrow}_{+}$ acts on the two-particle Hilbert space ${\cal H} \otimes {\cal H}$ in the following way:
\beq
U(g) \otimes U(g)~~|k\rangle \otimes |q\rangle = |g k\rangle \times |g q\rangle.
\eeq
There are similar equations for multiparticle states. 

Note that we can write $U(g) \otimes U(g) = [U \otimes U](g \times g)$.

Thus while defining the group action on multi-particle states, we see that we have made use of the isomorphism $G \rightarrow G \times G$ defined by $g \rightarrow g \times g$. This map is essential for the group action on multi-particle states. It is said to be a coproduct on $G$. We denote it by $\Delta$:
\beq
\Delta : G \rightarrow G \times G,
\eeq
\beq
\Delta(g) = g \times g.
\eeq

The coproduct exists in the algebra level also. Tensor products of representations of an algebra are in fact determined by $\Delta$ \cite{mack1, mack2}. It is a homomorphism from the group algebra $G^*$ to $G^* \otimes G^*$. A coproduct map need not be unique: Not all choices of $\Delta$ are equivalent. In particular the Clebsch-Gordan coefficients, which occur in the reduction of group representations, can depend upon $\Delta$. Examples of this sort occur for $\bar{P}^{\uparrow}_{+}$. In any case, it must fulfill
\bea
\label{eq:g1g2Action}
\Delta(g_{1})\Delta(g_{2}) &=& \Delta (g_{1}g_{2}), \; \; g_{1}, g_{2} \in G
\eea

Note that eqn. (\ref{eq:g1g2Action}) implies the coproduct on the group algebra $G^{*}$ by linearity. If $\alpha, \beta : G \rightarrow {\mathbb C}$ are smooth compactly supported functions on $G$, then the group algebra $G^{*}$ contains the generating elements
\beq
\int d\mu(g) \alpha(g) g,~~~~~~\int d\mu(g') \alpha(g') g',
\eeq
where $d\mu$ is the measure in $G$. The coproduct action on $G^{*}$ is then
\bea
\Delta : G^{*} &\rightarrow& G^{*} \otimes G^{*} \nn \\
\int d\mu(g) \alpha(g) g &\rightarrow&  \int d\mu(g) \alpha(g) \Delta(g).
\eea

The representations $U_{k}$ of $G^{*}$ on ${\cal H}_{k} (k = i, j)$,
\bea
U_{k} : \int d\mu(g) \alpha(g) g &\rightarrow&  \int d\mu(g) \alpha(g) U_{k}(g)
\eea
induced by those of $G$ also extend to the representation $U_{i} \otimes U_{j}$ on ${\cal H}_{i} \otimes {\cal H}_{j}$:
\bea
U_{i} \otimes U_{j} : \int d\mu(g) \alpha(g) g &\rightarrow&  \int d\mu(g) \alpha(g) (U_{i} \otimes U_{j})\Delta(g).
\eea

Thus the action of a symmetry group on the tensor product of representation spaces carrying any two representations $\rho_1$ and $\rho_2$ is determined by $\Delta$:
\beq
g \triangleright (\alpha \otimes \beta) = (\rho_1 \otimes \rho_2)\Delta(g)(\alpha \otimes \beta).
\eeq

If the representation space is itself an algebra ${\cal A}$, we have a rule for taking products of elements of ${\cal A}$ which involves the multiplication map $m$:
\bea
&& m : {\cal A} \otimes {\cal A} \rightarrow {\cal A},\\
&& \alpha \otimes \beta \rightarrow m(\alpha \otimes \beta)=\alpha \beta,
\eea
where $\alpha, \beta \in {\cal A}$.

It is now essential that $\Delta$ be compatible with $m$. That is 
\beq
\label{eq:compatibility}
m\Big[(\rho \otimes \rho)\Delta(g)(\alpha \otimes \beta)\Big]=\rho(g)m(\alpha \otimes \beta),
\eeq
where $\rho$ is a representation of the group acting on the algebra. 

The compatibility condition (\ref{eq:compatibility}) is encoded in the commutative diagram:
\begin{equation}
\begin{array}{ccc}
\alpha \otimes \beta & \longrightarrow & ( \rho \otimes \rho ) \Delta
(g) \alpha \otimes \beta \\
& &  \\ m \,\, \downarrow  & & \downarrow \,\, m \\ &  & \\ m(\alpha
\otimes
\beta) &
\longrightarrow & \rho(g) m (\alpha \otimes \beta)
\end{array}
\end{equation}
If such a $\Delta$ can be found, $G$ is an automorphism of ${\cal A}$. In the absence of such a $\Delta$, $G$ does not act on ${\cal A}$.

Let us consider the action of $P_{+}^{\uparrow}$ on the nocommutative spacetime algebra (GM plane)  ${\cal A}_{\theta}({\mathbb R}^{d+1})$. The algebra ${\cal A}_{\theta}({\mathbb R}^{d+1})$ consists of smooth functions on ${\mathbb R}^{d+1}$ with the multiplication map
\beq
m_{\theta}: {\cal A}_{\theta}({\mathbb R}^{d+1}) \otimes {\cal A}_{\theta}({\mathbb R}^{d+1}) \rightarrow {\cal A}_{\theta}({\mathbb R}^{d+1}).
\eeq
For two functions $\alpha$ and $\beta$ in the algebra ${\cal A}_{\theta}$, the multiplication map is not a point-wise multiplication, it is the $*$-multiplication:
\beq
m_{\theta} (\alpha \otimes \beta) (x) = (\alpha * \beta)(x). 
\eeq
Explicitly the $*$-product between two functions $\alpha$ and $\beta$ is written as
\beq
(\alpha * \beta)(x) = \textrm{exp}\Big(\frac{i}{2}\theta^{\mu \nu}\frac{\partial}{\partial x^{\mu}}\frac{\partial}{\partial y^{\nu}}\Big)\alpha(x)\beta(y)\Big|_{x=y}.
\eeq

Before implementing the Poincar\'e group action on ${\cal A}_{\theta}$, we write down a useful expression for $m_{\theta}$ in terms of the commutative multiplication map $m_{0}$,
\beq
m_{\theta} = m_{0} {\cal F}_{\theta},
\eeq
where 
\beq
{\cal F}_{\theta} = \textrm{exp}(-\frac{i}{2}\theta^{\alpha \beta}P_{\alpha} \otimes P_{\beta}), \; \; \; P_{\alpha} = -i \partial_{\alpha}
\eeq
is called the ``Drinfel'd twist" or simply the ``twist". The indices here are raised or lowered with the Minkowski metric with signature ($+, -, -, -$).

It is easy to show from this equation that the Poincar\'e group action through the coproduct $\Delta(g)$ on the noncommutative algebra of functions is not compatible with the $*$-product. That is, $P_{+}^{\uparrow}$ does not act on ${\cal A}_{\theta}({\mathbb R}^{d+1})$ in the usual way. There is a way to implement Poincar\'e symmetry on noncommuative algebra. Using the twist element, the coproduct of the universal enveloping algebra ${\cal U}({\cal P})$ of the Poincar\'e algebra can be deformed in such a way that it is compatible with the above $*$-multiplication. The deformed coproduct, denoted by $\Delta_{\theta}$ is:
\beq
\Delta_{\theta} = {\cal F}^{-1}_{\theta} \Delta {\cal F}_{\theta}
\eeq

We can check compatibility of the twisted coproduct $\Delta_{\theta}$ with the twisted multiplication $m_{\theta}$ as follows
\begin{eqnarray}
m_{\theta} \left( (\rho \otimes \rho) \Delta_{\theta}(g) ( \alpha \otimes
\beta ) \right) &=&
m_{0} \left( {\cal F}_{\theta} ({\cal F}_{\theta}^{-1} \rho(g) \otimes
\rho(g) {\cal F}_{\theta}) \alpha \otimes \beta \right) \nonumber \\
&=& \rho(g) \left( \alpha * \beta \right), \quad
 \alpha,\beta \in {\cal A}_\theta(\rr^{d+1})
\label{proofcomp}
\end{eqnarray}
as required.
This compatibility is encoded in the commutative diagram
\begin{equation}
\begin{array}{ccc}
\alpha \otimes \beta & \longrightarrow & ( \rho \otimes \rho ) \Delta_{\theta}
(g) \alpha \otimes \beta \\
& &  \\ m_{\theta} \,\, \downarrow  & & \downarrow \,\, m_{\theta} \\ &  & \\ \alpha * \beta &
\longrightarrow & \rho(g) (\alpha * \beta)
\end{array}
\end{equation}
Thus $G$ is an automorphism of ${\cal A}_{\theta}$ if the coproduct is $\Delta_{\theta}$.
 
It is easy to see that the coproduct for the generators $P_{\alpha}$ of the Lie algebra of the translation group are not deformed, 
\beq
\Delta_{\theta}(P_{\alpha}) = \Delta(P_{\alpha})
\eeq
while the coproduct for the generators of the Lie algebra of the Lorentz group are deformed:
\bea
\Delta_{\theta}(M_{\mu \nu}) &=& 1 \otimes M_{\mu \nu} + M_{\mu \nu} \otimes 1 - \frac{1}{2}\Big[(P\cdot \theta)_{\mu}\otimes P_{\nu}-P_{\nu}\otimes (P\cdot \theta)_{\mu} - (\mu \leftrightarrow \nu) \Big], \nn \\
(P \cdot \theta)_{\lambda} &=& P_{\rho}\theta^{\rho}_{\lambda}.
\eea

The idea of twisting the coproduct in noncommutative spacetime algebra is due to \cite{chaichian, drinfeld, majid, fiore2, fiore1, fioresolo1, fioresolo2, watts1, Oeckl:2000eg, watts2, gms, Dimitrijevic:2004rf, matlock, aschieri3}. But its origins can be traced back to Drinfel'd \cite{drinfeld} in mathematics. This Drinfel'd twist leads naturally to deformed $R$-matrices and statistics for quantum groups, as discussed by Majid \cite{majid}. Subsequently, Fiore and Schupp \cite{fiore1} and Watts \cite{watts1,watts2} explored the significance of the Drinfel'd twist and $R$-matrices while Fiore \cite{fioresolo1, fioresolo2} and Fiore and Schupp \cite{fiore2}, Oeckl \cite{Oeckl:2000eg} and Grosse {\it et al.} \cite{gms} studied the importance of $R$-matrices for statistics. Oeckl \cite{Oeckl:2000eg} and Grosse {\it et al.} \cite{gms} also developed quantum field theories using different and apparently inequivalent approaches, the first on the Moyal plane and the second on the $q$-deformed fuzzy sphere. In \cite{aschieri3, Dimitrijevic:2004rf} the authors focused on the diffiomorphism group $\mathcal{D}$ and developed Riemannian geometry and gravity theories based on $\Delta_\theta$, while \cite{chaichian} focused on the Poincar\'{e} subgroup $\mathcal{P}$ of $\mathcal{D}$ and explored the consequences of $\Delta_\theta$ for quantum field theories. Twisted conformal symmetry was discussed by \cite{matlock}. Recent work, including ours \cite{bal-unitary, bal, uv-ir, bal-sasha-babar, bal-stat, cpt-paper, twistd}, has significant overlap with the earlier literature.
\subsection{The Twisted Statistics}
In the previous section, we discussed how to implement the Poincar\'e group action in the noncommutative framework. We changed the ordinary coproduct to a twisted coproduct $\Delta_{\theta}$ to make it compatible with the multiplication map $m_{\theta}$. This very process of twisting the coproduct has an impact on statistics. In this section we discuss how the deformed Poincar\'e symmetry leads to a new kind of statistics for the particles.

Consider a two-particle system in quantum mechanics for the case $\theta^{\mu \nu}=0$. A two-particle wave function is a function of two sets variables, and lives in ${\cal A}_{0} \otimes {\cal A}_{0}$. It transforms according to the usual coproduct $\Delta$. Similarly in the noncommutative case, the two-particle wave function lives in ${\cal A}_{\theta} \otimes {\cal A}_{\theta}$ and transforms according to the twisted coproduct $\Delta_{\theta}$.

In the commutative case, we require that the physical wave functions describing identical particles are either symmetric (bosons) or antisymmetric (fermions), that is, we work with either the symmetrized or antisymmetrized tensor product,
\bea
\phi \otimes_{S} \chi &\equiv& \frac{1}{2}\left(\phi \otimes \chi 
+ \chi
\otimes \phi \right),\\
\phi \otimes_{A} \chi &\equiv& \frac{1}{2}\left(\phi \otimes \chi 
- \chi
\otimes \phi \right).
\eea
which satisfies
\bea
\phi \otimes_{S} \chi &=& + \chi
\otimes_{S} \phi,\\
\phi \otimes_{A} \chi &=& - \chi
\otimes_{A} \phi.
\eea
These relations have to hold in all frames of reference in a Lorentz-invariant theory. That is, symmetrization and antisymmetrization must commute with the Lorentz group action. 

Since $\Delta(g)=g \times g$, we have
\beq
\tau_0 (\rho \otimes \rho) \Delta(g)=(\rho \times \rho)\Delta(g) \tau_0, \; \; g \in P_{+}^{\uparrow}
\eeq
where $\tau_{0}$ is the flip operator:
\beq
\tau_{0} (\phi \otimes \chi) = \chi \otimes \phi.
\eeq

Since 
\beq
\phi \otimes_{S, A} \chi =\frac{1 \pm \tau_{0}}{2}~\phi \otimes \chi,
\eeq
we see that Lorentz transformations preserve symmetrization and anti-symmetrization. 

The twisted coproduct action of the Lorentz group is not compatible with the usual symmetrization and anti-symmetrization. The origin of this fact can be traced to the fact that the coproduct is not cocommutative except when $\theta^{\mu \nu}={0}$. That is,
\bea
\tau_{0}{\cal F}_{\theta} &=& {\cal F}_{\theta}^{-1}\tau_{0}, \\
\tau_{0} (\rho \otimes \rho) \Delta_{\theta}(g) &=& (\rho \otimes \rho)\Delta_{-\theta}(g)\tau_{0}
\eea

One can easily construct an appropriate deformation $\tau_{\theta}$ of the operator $\tau_{0}$ using the twist operator ${\cal F}_{\theta}$ and the definition of the twisted coproduct, such that it commutes with $\Delta_{\theta}$. Since $\Delta_{\theta}(g)={\cal F}_{\theta}^{-1}\Delta(g){\cal F}_{\theta}$, it is
\bea
\tau_{\theta} &=& {\cal F}_{\theta}^{-1} \tau_{0} {\cal F}_{\theta}.
\eea
It has the property,
\bea
(\tau_\theta)^2 &=& {\bf 1}\otimes {\bf 1}.
\eea

The states constructed according to
\beq
\phi \otimes_{S_\theta} \chi \equiv
\left(\frac{1\,+ \tau_\theta}{2}\right)\,
(\phi\,\otimes\,\chi),
\eeq
\beq
\phi \otimes_{A_\theta} \chi \equiv \\
\left(\frac{1\,-\tau_\theta}{2}\right)\,(\phi\,\otimes\,\chi)
\eeq
form the physical two-particle Hilbert spaces of (generalized) bosons and fermions obeying twisted statistics.
\subsection{Statistics of Quantum Fields}
The very act of implementing Poincar\'e symmetry on a noncommutative spacetime algebra leads to twisted fermions and bosons. In this section we look at the second quantized version of the theory and we encounter another surprise on the way.

We can connect an operator in Hilbert space and a quantum field in the following way. A quantum field on evaluation at a spacetime point gives an operator-valued distribution acting on a Hilbert space. A quantum field at a spacetime point $x_{1}$ acting on the vacuum gives a one-particle state centered at $x_{1}$. Similarly we can construct a two-particle state in the Hilbert space. The product of two quantum fields at spacetime points $x_{1}$ and $x_{2}$ when acting on the vacuum generates a two-particle state where one particle is centered at $x_{1}$ and the other at $x_{2}$.

In the commutative case, a free spin-zero quantum scalar field $\varphi_{0}(x)$ of mass $m$ has the mode expansion
\beq
\varphi_{0}(x) =\int d \mu(p) \; (c_{\bf p}\; \textrm{e}_{p}(x) + d_{\bf p}^{\dagger} \; \textrm{e}_{-p}(x))
\eeq
where
\beq
\textrm{e}_{p}(x) = \textrm{e}^{-i\; p\cdot x}, \; \; p \cdot x = p_{0}x_{0} - {\bf p}\cdot {\bf x}, \; \; d \mu(p) = \frac{1}{(2\pi)^{3}}\frac{d^{3}p}{2p_{0}}, \; \; \; p_{0} = \sqrt{{\bf p}^{2} + m^{2}} > 0. \nn
\eeq

The annihilation-creation operators $c_{{\bf p}}$, $c^{\dagger}_{{\bf p}}$, $d_{{\bf p}}$,
$d_{{\bf p}}^{\dagger}$ satisfy the standard commutation relations,
\bea
\label{eq:standard}
c_{{\bf p}}c_{{\bf q}}^{\dagger} \pm c_{{\bf q}}^{\dagger}c_{{\bf p}} &=& 2p_{0} \;\delta^{3}({\bf p}-{\bf q})\\
d_{{\bf p}}d_{{\bf q}}^{\dagger} \pm d_{{\bf q}}^{\dagger}d_{{\bf p}} &=& 2p_{0} \;\delta^{3}({\bf p}-{\bf q}).
\eea
The remaining commutators involving these operators vanish. 

If $c_{\bf p}$ is the annihilation operator of the second-quantized field $\varphi_{0}(x)$, an elementary calculation tells us that
\bea
\label{eq:qft-qm}
\langle 0 |\varphi_{0}(x) c^\dagger_{\bf p} |0\rangle &=& e_p(x)= 
e^{-i p \cdot x}.\nonumber
\eea
\bea
\frac{1}{2}\langle 0 |\varphi_{0}(x_1) \varphi_{0}(x_2)
c^\dagger_{\bf q}
c^\dagger_{\bf p} |0\rangle &=&  \nonumber
\left(\frac{{\bf 1} \pm
\tau_{0}}{2}\right)(e_p \otimes e_q)(x_1,x_2) \\
&\equiv& (e_p \otimes_{S_0,A_0} e_q)(x_1,x_2)\nn\\
&\equiv& \langle x_{1}, x_{2}|p, q\rangle_{S_0,A_0}.
\eea
where we have used the commutation relation
\beq
c_{\bf p}^\dagger~c_{\bf q}^\dagger\,= \pm~c_{\bf q}^\dagger~c_{\bf p}^\dagger~.
\eeq

From the previous section we have learned that the two-particle states in noncommutative spacetime should be constructed in such a way that they obey twisted symmetry. That is,
\beq
|p, q\rangle_{S_0,A_0}~\rightarrow~|p, q\rangle_{S_\theta,A_\theta}.
\eeq 
This can happen only if we modify the quantum field $\varphi_{0}(x)$ in such a way that the analogue of eqn. (\ref{eq:qft-qm}) in the noncommutative framework gives us $|p, q\rangle_{S_\theta,A_\theta}$. Let us denote the modified quantum field by $\varphi_{\theta}$. It has a mode expansion
\beq
\varphi_{\theta}(x) =\int d \mu(p) \; (a_{\bf p}\; \textrm{e}_{p}(x) + b_{\bf p}^{\dagger} \; \textrm{e}_{-p}(x))
\eeq

Noncommutativity of spacetime does not change the dispersion relation for the quantum field in our framework. It will definitely change the operator coefficients of the plane wave basis. Here we denote the new $\theta$-deformed annihilation-creation operators by $a_{{\bf p}}$, $a^{\dagger}_{{\bf p}}$, $b_{{\bf p}}$, $b_{{\bf p}}^{\dagger}$. Let us try to connect the quantum field in noncommutative spacetime with its counterpart in commutative spacetime, keeping in mind that they should coincide in the limit $\theta^{\mu \nu} \rightarrow 0$.

The two-particle state $|p, q\rangle_{S_{\theta}, A_{\theta}}$ for bosons and fermions obeying deformed statistics is constructed as follows:
\bea
|p, q\rangle_{S_{\theta}, A_{\theta}} &\equiv& |p\rangle \otimes_{_{S_{\theta}, A_{\theta}}} |q\rangle =\Big(\frac{1 \pm \tau_{\theta}}{2}\Big) (|p\rangle \otimes |q\rangle)\nn \\
&=& \frac{1}{2}\Big(|p\rangle \otimes |q\rangle \pm \textrm{e}^{-i q_{\mu}\theta^{\mu \nu}p_{\nu}}|q\rangle \otimes |p\rangle\Big).
\eea

Exchanging $p$ and $q$ in the above, one finds
\beq
\label{eq:pq-qp}
|p, q\rangle_{S_{\theta}, A_{\theta}} = \pm \; \textrm{e}^{i p_{\mu}\theta^{\mu \nu}q_{\nu}}|q, p\rangle_{S_{\theta}, A_{\theta}}.
\eeq

In Fock space the above two-particle state is constructed from the modified second-quantized field $\varphi_{\theta}$ according to
\bea
\frac{1}{2}\langle0|\varphi_{\theta}(x_{1})\varphi_{\theta}(x_{2}) a_{\bf q}^{\dagger}a_{\bf p}^{\dagger}|0\rangle &=& \Big(\frac{1\pm \tau_{\theta}}{2}\Big) (e_{p} \otimes e_{q})(x_{1}, x_{2})\nn \\
&=& (e_{p} \otimes_{S_{\theta}, A_{\theta}} e_{q})(x_{1}, x_{2})\nn \\
&=& \langle x_1, x_2|p, q\rangle_{S_{\theta}, A_{\theta}}.
\eea

On using eqn. (\ref{eq:pq-qp}), this leads to the relation 
\beq
\label{non-commu-1}
a_{\bf p}^{\dagger}a_{\bf q}^{\dagger} =  \pm \; \textrm{e}^{i p_{\mu}\theta^{\mu \nu}q_{\nu}}\; a_{\bf q}^{\dagger}a_{\bf p}^{\dagger}.
\eeq

It implies
\bea
\label{non-commu-2}
a_{\bf p}a_{\bf q} &=&  \pm \; \textrm{e}^{i p_{\mu}\theta^{\mu \nu}q_{\nu}}\; a_{\bf q}a_{\bf p.}\eea

Thus we have a new type of bilinear relations reflecting the deformed quantum symmetry. 

This result shows that while constructing a quantum field theory on noncommutative spacetime, we should twist the creation and annihilation operators in addition to the $*$-multiplication between the fields.

In the limit $\theta^{\mu \nu}=0$, the twisted creation and annihilation operators should match with their counterparts in commutative case. There is a way to connect these operators in the two cases. The transformation connecting the twisted operators,  $a_{{\bf p}}$, $b_{{\bf p}}$, and the untwisted operators, $c_{{\bf p}}$, $d_{{\bf p}}$, is called the ``dressing transformation" \cite{Grosse, Faddeev-Zamolodchikov}. It is defined as follows:
\beq
\label{dressingT}
a_{{\bf p}} = c_{{\bf p}} \; e^{-\frac{i}{2}p_{\mu} \theta^{\mu\nu}P_{\nu}},\; \; \;
b_{{\bf p}} = d_{{\bf p}}\; e^{-\frac{i}{2}p_{\mu}\theta^{\mu\nu}P_{\nu}},
\eeq
where $P_{\mu}$ is the four-momentum operator,
\beq
P_{\mu} = \int \frac{d^{3}p}{2p_{0}}\; (c_{{\bf p}}^\dagger c_{{\bf p}} +
d^{\dagger}_{{\bf p}}
d_{{\bf p}})\; p_{\mu}.
\eeq

The Grosse-Faddeev-Zamolodchikov algebra is the above twisted or dressed algebra \cite{Grosse, Faddeev-Zamolodchikov}. (See also \cite{queiroz1, queiroz2} in this connection.)

Note that the four-momentum operator $P_{\mu}$ can also be written in terms of the twisted operators:
\beq
\label{eq:pmu1}
P_{\mu} =\int \frac{d^{3}p}{2p_{0}}\; (a_{{\bf p}}^{\dagger} a_{{\bf p}} +
b^{\dagger}_{{\bf p}}
b_{{\bf p}}) \; p_{\mu}.
\eeq
That is because $p_{\mu} \theta^{\mu\nu}P_{\nu}$ commutes with any of the operators for momentum $p$. For example 
\beq
[P_{\mu},a_{{\bf p}}]=-p_{\mu} a_{{\bf
p}},
\eeq
so that 
\beq
[p_{\nu} \theta^{\nu\mu}P_{\mu}, a_{{\bf p}}]= p_{\nu}\theta^{\nu\mu}p_{\mu}=0,
\eeq
$\theta$ being antisymmetric. 

The antisymmetry of $\theta^{\mu \nu}$ allows us to write
\beq
c_{\bf p}e^{-\frac{i}{2}p_{\mu} \theta^{\mu \nu} P_{\nu}} = e^{-\frac{i}{2}p_{\mu}
\theta^{\mu \nu} P_{\nu}} c_{\bf p}, 
\eeq
\beq
c^{\dagger}_{\bf p}e^{\frac{i}{2}p_{\mu} \theta^{\mu \nu} P_{\nu}} = e^{\frac{i}{2}p_{\mu} \theta^{\mu \nu} P_{\nu}}c^{\dagger}_{\bf p}. 
\eeq
Hence the ordering of factors here is immeterial. 

It should also be noted that the map from the $c$- to the $a$-operators is invertible,
\beq
c_{\bf p} = a_{\bf p} \; e^{\frac{i}{2}p_{\mu} \theta^{\mu\nu}P_{\nu}},\; \; \;
d_{\bf p} =
b_{\bf p}\; e^{\frac{i}{2}p_{\mu}\theta^{\mu\nu}P_{\nu}},\nn
\eeq
where $P_{\mu}$ is written as in eqn.~(\ref{eq:pmu1}). 

The $\star$-product between the modified (twisted) quantum fields is
\beq
(\varphi_{\theta} \star \varphi_{\theta})(x) = \varphi_{\theta}(x) e^{\frac{i}{2} \overleftarrow{\partial}
\wedge \overrightarrow{\partial}} \varphi_{\theta}(y) |_{x=y},
\eeq
\beq
\overleftarrow{\partial} \wedge \overrightarrow{\partial} :=
\overleftarrow{\partial}_{\mu} \theta^{\mu \nu} \overrightarrow{\partial}_{\nu}.\nn
\eeq

The twisted quantum field $\varphi_{\theta}$ differs from the untwisted quantum field
$\varphi_{0}$ in two ways:
\begin{center}
$i.)$ $e_{p} \in {\cal A}_\theta(\rr^{d+1})~~~~~~~~~~~~~~~~~~~~~~~~~~~~~~~~~~~~~~~~~~~~~~~~$
\end{center}
and
\begin{center}
$ii.)$ $a_{{\bf p}}$ is twisted by statistics. $~~~~~~~~~~~~~~~~~~~~~~~~~~~~~~~~$
\end{center}
The twisted statistics can be accounted by writing \cite{bal-sasha-babar}
\beq
\varphi_{\theta} = \varphi_{0} \; e^{\frac{i}{2} \overleftarrow{\partial} \wedge P},
\label{eq:gaugefield}
\eeq
where $P_{\mu}$ is the total momentum operator. From this follows that the
$\star$-product of an
arbitrary number of fields $\varphi_{\theta}^{(i)}$ ($i$ = 1, 2, 3, $\cdots$) is
\beq
\varphi_{\theta}^{(1)} \star \varphi_{\theta}^{(2)} \star {\cdots} =(\varphi^{(1)}_{0}\varphi^{(2)}_{0} {\cdots}) \; e^{\frac{i}{2} \overleftarrow{\partial} \wedge P}.
\label{eq:productfields}
\eeq

Similar deformations occur for all tensorial and spinorial quantum fields.

In \cite{cmb}, a noncommutative cosmic microwave background (CMB) power spectrum is calculated by promoting the quantum fluctuations $\varphi_{0}$ of the scalar field driving inflation (the inflaton) to a twisted quantum field $\varphi_{\theta}$. The power spectrum becomes direction-dependent, breaking the statistical anisotropy of the CMB. Also, $n$-point correlation functions become non-Gaussian when the fields are noncommutative, assuming that they are Gaussian in their commutative limits. These effects can be tested experimentally.

In this article we discuss field theory with spacetime noncommutativity. It should also be noted that there is another approach in which noncommutativity is encoded in the degrees of freedom of the fields while keeping spacetime commutative \cite{Carmona, Carmona1}. Such noncommutativity can also be interpreted in terms of twisted statistics. In \cite{queiroz1} a noncommutative black body spectrum is calculated using this approach (which is based on \cite{Carmona, Carmona1}). Also, a noncommutative-gas driven inflation is considered in \cite{queiroz2} along this formulation.

\subsection{From Twisted Statistics to Noncommutative Spacetime}
Noncommutative spacetime leads to twisted statistics. It is also possible to start from a twisted statistics and end up with a noncommutative spacetime \cite{queiroz, Bal-Queiroz}. Consider the commutative version $\varphi_{0}$ of the above quantum field $\varphi_{\theta}$. The creation and annihilation operators of this field fulfill the standard commutation relations as given in eqn. (\ref{eq:standard}).

Let us twist statistics by deforming the creation-annihilation operators $c_{\bf p}$ and $c_{\bf p}^\dagger$ to
\beq
a_{\bf p} =c_{\bf p}\;e^{-\frac{i}{2} \; p_\mu \;\theta^{\mu \nu}\;P_\nu}\;, \hspace{10mm}
a^\dagger_{\bf p} = c^\dagger_{\bf p}\;e^{\frac{i}{2} \; p_\mu \;\theta^{\mu \nu}\;P_\nu} 
\eeq

Now statistics is twisted since $a$'s and $a^\dagger$'s no longer fulfill standard relations. They obey the relations given in eqn. (\ref{non-commu-1}) and eqn. (\ref{non-commu-2}) This twist affects the usual symmetry of particle interchange. The $n$-particle wave function $\psi_{k_1 \cdots k_n}$,
\beq
\psi_{k_1, \cdots, k_n}(x_1, \ldots, x_n) = \langle 0|\varphi(x_1)\varphi(x_2)\ldots \varphi(x_n)~a^\dagger_{{\bf k}_n}a^\dagger_{{\bf k}_{n-1}}\ldots
a^\dagger_{{\bf k}_1}\;|0\rangle 
\eeq
is no longer symmetric under the interchange of $k_i$. It fullfils a twisted symmetry given by
\beq
\label{eq:42.0}
\psi_{k_1 \cdots k_i \; k_{i+1} \cdots k_n} = \textrm{exp}\Big(-i k_{i}^{\mu} \; \theta_{\mu \nu} \; k_{i+1}^{\nu}\Big) \; \psi_{k_1 \cdots k_{i+1} \; k_i \cdots k_n } 
\eeq
showing that statistics is twisted. We can show that this in fact leads to a noncommutative spacetime if we require Poincar\'e invariance. It is explained below.

In the commutative case, the elements $g$ of $P_{+}^{\uparrow}$ acts on $\psi_{k_1 \cdots k_n}$ by the representative $U(g)\otimes U(g) \otimes \cdots \otimes U(g)$ ($n$ factors) compatibly with the symmetry of $\psi_{k_1 \cdots k_n}$. This action is based on the coproduct
\beq
\Delta(g) = g \times g\;.
\eeq

But for $\theta^{\mu \nu} \neq 0$, and for $g\neq\textrm{identity}$, already for the case $n=2$,
\bea
\Delta(g) \psi_{p, q} &=& \psi_{gp, gq}\nn \\
&=& e^{-i p_{\mu} \theta^{\mu \nu} q_{\nu}} \Delta(g) \psi_{q, p}\nn \\
&=& e^{-i p_{\mu} \theta^{\mu \nu} q_{\nu}} \psi_{gq, gp}\nn \\
&\neq& e^{-i (gp)_{\mu}\theta^{\mu \nu}(gq)_{\nu}} \psi_{gq, gp}.
\eea

Thus the usual coproduct $\Delta_0$ is not compatible with the statistics (\ref{eq:42.0}). It has to be twisted to
\beq
\label{eq:43}
\Delta_\theta(g) = {\cal F}^{-1}_\theta \Delta (g) {\cal F}_\theta,~~\Delta(g)=(g\times g)
\eeq
to be compatible with the new statistics. At this point $\Delta_\theta(g)$ is not compatible with $m_{0}$, the commutative (point-wise) multiplication map. So we are forced to change the multiplication map to $m_{\theta}$,
\beq
m_{\theta} = m_{0} \; {\cal F}_\theta
\eeq
for this compatibility. Since
\beq
m_\theta(\alpha \otimes \beta) = \alpha * \beta,
\eeq
we end up with noncommutative spacetime. Thus twisted statistics can lead to spacetime noncommutativity.
\subsection{Violation of the Pauli Principle}
In section 4.3, we wrote down the twisted commutation relations. In the fermionic sector, these relations read
\bea
\label{phaseFermion1}
a_{\bf p}^{\dagger}a_{\bf q}^{\dagger} + \; \textrm{e}^{i p_{\mu}\theta^{\mu \nu}q_{\nu}}\; a_{\bf q}^{\dagger}a_{\bf p}^{\dagger}&=&0 \\
\label{phaseFermion2}
a_{\bf p}a_{\bf q}^{\dagger} + \; \textrm{e}^{-i p_{\mu}\theta^{\mu \nu}q_{\nu}}\; a_{\bf q}^{\dagger}a_{\bf p} &=& 2q_{0}\delta^{3}({\bf p} - {\bf q}).
\eea

In the commutative case, above relations read
\bea
c_{\bf p}^{\dagger}c_{\bf q}^{\dagger} + c_{\bf q}^{\dagger}c_{\bf p}^{\dagger}&=&0 \\
c_{\bf p}c_{\bf q}^{\dagger} + c_{\bf q}^{\dagger}c_{\bf p} &=& 2q_{0}\delta^{3}({\bf p} - {\bf q}).
\eea
The phase factor appearing in eqn (\ref{phaseFermion1}) and eqn. (\ref{phaseFermion2}) while exchanging the operators has a nontrivial physical consequence which forces us to reconsider the Pauli exclusion principle. A modification of Pauli principle compatible with the twisted statistics can lead to Pauli forbidden processess and they can be subjected to stringent experimental tests. 

For example, there are results from SuperKamiokande \cite{sk} and Borexino \cite{borexino} putting limits on the violation of Pauli exclusion principle in nucleon systems. These results are based on non-observed transition from Pauli-allowed states to Pauli-forbidden states with $\beta^{\pm}$ decays or $\gamma$, $p$, $n$ emission. A bound for $\theta$ as strong as $10^{11}$ Gev is obtained from these results \cite{Gianpiero}. 
\subsection{Statisitcal Potential}
Twisting the statistics can modify the spatial correlation functions of fermions and bosons and thus affect the statistical potential existing between any two particles.

Consider a canonical ensemble, a system of $N$ indistinguishable, non-interacting particles confined to a three-dimensional cubical box of volume $V$, characterized by the inverse temperature $\beta$. In the coordinate representation, we write down the density matrix of the system \cite{pathria}
\beq
\langle {\bf r}_{1}, \cdots {\bf r}_{N}|\hat{\rho}|{\bf r}'_{1}, \cdots {\bf r}'_{N}\rangle = \frac{1}{Q_{N}(\beta)}\langle {\bf r}_{1}, \cdots {\bf r}_{N}|\textrm{e}^{-\beta \hat{H}}|{\bf r}'_{1}, \cdots {\bf r}'_{N}\rangle,
\eeq
where $Q_{N}(\beta)$ is the partition function of the system given by
\beq
Q_{N}(\beta) = \textrm{Tr}(\textrm{e}^{-\beta \hat{H}}) = \int d^{3N} r \langle {\bf r}_{1}, \cdots {\bf r}_{N}|\textrm{e}^{-\beta \hat{H}}|{\bf r}'_{1}, \cdots {\bf r}'_{N}\rangle .
\eeq

Since the particles are non-interacting, we may write down the eigenfunctions and eigenvalues of the system in terms of the single-particle wave functions and single-particle energies.

For free non-relativistic particles, we have the energy eigenvalues
\beq
E = \frac{\hbar^{2}}{2m}\sum_{i=1}^{N} k_{i}^{2}
\eeq
where $k_{i}$ is the magnitude of the wave vector of the $i$-th particle. Imposing periodic boundary conditions, we write down the normalized single-particle wave function
\beq
u_{{\bf k}}({\bf r}) = V^{-1/2} \textrm{e}^{i {\bf k} \cdot {\bf r}}
\eeq
with ${\bf k} = 2 \pi V^{-1/3} {\bf n}$ and ${\bf n}$ is a three-dimensional vector whose components take values $0, \pm 1, \pm 2, \cdots$.

Following the steps given in \cite{pathria}, we write down the diagonal elements of the density matrix for the simplest relevant case with $N=2$,
\beq
\label{approxTwoPart}
\langle {\bf r}_{1}, {\bf r}_{2}|\hat{\rho}|{\bf r}_{1}, {\bf r}_{2}\rangle \approx \frac{1}{V^{2}} \big(1 \pm \textrm{exp}(-2 \pi r_{12}^{2}/\lambda^{2})\Big)
\eeq
where the plus and the minus signs indicate bosons and fermions respectively, $r_{12}=|{\bf r}_{1} - {\bf r}_{2}|$ and $\lambda$ is the mean thermal wavelength,
\beq
\lambda = \hbar \sqrt{\frac{2 \pi \beta}{m}},~~~~~~\beta = \frac{1}{k_{B}T}.
\eeq

Note that eqn. (\ref{approxTwoPart}) is obtained under the assumption that the mean interparticle distance $(V/N)^{1/3}$ in the system is much larger than the mean thermal wavelength $\lambda$. Eqn. (\ref{approxTwoPart}) indicates that spatial correlations are non-zero even when the particles are non-interacting. These correlations are purely due to statistics: They emerge from the symmetrization or anti-symmetrization of the wave functions describing the particles. Particles obeying Bose statistics give a positive spatial correlation and particles obeying Fermi statistics give a negative spatial correlation.  

We can express spatial correlations between particles by introducing a statistical potential $v_{s}(r)$ and thus treat the particles classically \cite{Uhlenbeck}. The statistical potential corresponding to the spatial correlation given in eqn. (\ref{approxTwoPart}) is
\beq
v_{s}(r) = -k_{B}T \; \textrm{ln} \Big(1 \pm \textrm{exp}(-2\pi r_{12}^{2}/\lambda^{2})\Big)
\eeq

From this equation, it follows that two bosons always experience a ``statistical attraction" while two fermions always experience a ``statistical repulsion". In both cases, the potential decays rapidly when $r > \lambda$.

So far our discussion focussed on particles in commutative spacetime. We can derive an expression for the statistical potential between two particles living in a noncommutative spacetime. The results \cite{correlation} are interesting. In a noncommutative spacetime with 2+1 dimensions and for the case $\theta^{0i} = 0$, we write down the answer for the spatial correlation between two non-interacting particles from \cite{correlation}
\beq
\label{two-P-theta}
\langle {\bf r}_{1}, {\bf r}_{2}|\hat{\rho}|{\bf r}_{1}, {\bf r}_{2}\rangle_{\theta} \approx \frac{1}{A^2}\left(1\pm\frac{1}{1+\frac{\theta^2}{\lambda^4}}
e^{- 2 \pi\, r_{12}^2/(\lambda^{2}(1+\frac{\theta^2}{\lambda^4}))}\right)
\eeq
Here $A$ is the area of the system. This result can be generalized to higher dimensions by replacing $\theta^2$ by an appropriate sum of $(\theta^{ij})^2$ \cite{correlation}. It reduces to the standard (untwisted) result given in eqn. (\ref{approxTwoPart}) in the limit $\theta\rightarrow0$.
\begin{figure}
\centerline{\epsfig{figure=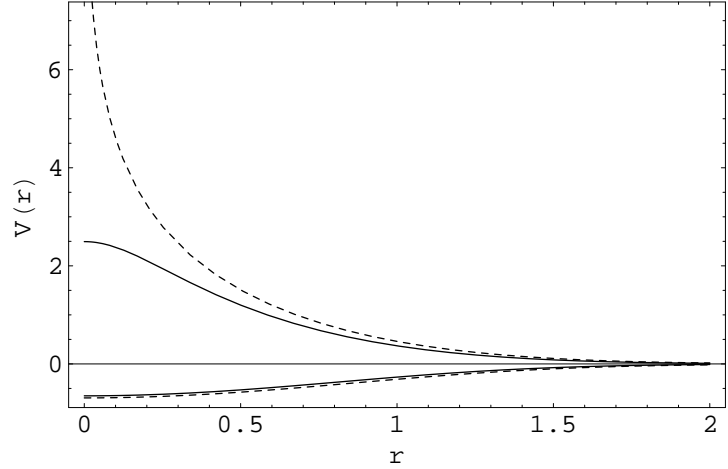}}
\caption {Statistical potential $v(r)$ measured in units of $k_BT$.  An irrelevant additive constant has been set zero.  The upper two curves represent the fermionic cases and the lower curves the bosonic cases.  The solid line shows the noncommutative result and the dashed line the commutative case. The curves are drawn for the value $\frac{\theta}{\lambda^2}=0.3$. The separation $r$ is measured in units of the thermal length $\lambda$. \cite{correlation}}
\label{fig:potential}
\end{figure}
Notice that the spatial correlation function for fermions does not vanish in the limit $r\rightarrow 0$ (See Fig. \ref{fig:potential}).  That means that there is a finite probability that fermions may come very close to each other. This probability is determined by the noncommutativity parameter $\theta$. Also notice that the assumptions made in \cite{correlation} are valid for low temperature and low density limits. At high temperature and high density limits a much more careful analysis is required to investigate the noncommutative effects.
\section{Matter Fields, Gauge Fields and Interactions}
In section 4, we discussed the statistics of quantum fields by taking a simple example of a massive, spin-zero quantum field. In this section, we discuss how matter and gauge fields are constructed in the noncommutative formulation and their interactions. We also explain some interesting results which can be verified experimentally.
\subsection{Pure Matter Fields}
Consider a second quantized real Hermitian field of mass $m$,
\beq
\Phi = \Phi^{-} + \Phi^{+}
\eeq
where the creation and annihilation fields are constructed from the creation and annihilation operators:
\bea
\Phi^{-}(x) &=& \int d\mu(p) \; e^{ipx} \; a_{\bf p}^{\dagger}\\
\Phi^{+}(x) &=& \int d\mu(p) \; e^{-ipx} \; a_{\bf p}
\eea
The deformed quantum field $\Phi$ can be written in terms of the un-deformed quantum field $\Phi_{0}$,
\beq
\label{Theta-to-zero}
\Phi(x) = \Phi_{0}(x) e^{\frac{1}{2}\overleftarrow{\partial}^{\mu}\theta_{\mu \nu}P^{\nu}}
\eeq
where the creation and annihilation fields of the un-deformed quantum field is constructed from the usual creation and annihilation operators
\bea
\Phi_{0}^{-}(x) &=& \int d\mu(p) \; e^{ipx} \; c_{\bf p}^{\dagger}, \\
\Phi_{0}^{+}(x) &=& \int d\mu(p) \; e^{-ipx} \; c_{\bf p}
\eea

When evaluating the product of $\Phi$'s at the same point, we must take $*$-product of the $e_{p}$'s since $e_{p} \in {\cal A}_{\theta}({\mathbb R}^{N})$. We can make use of eqn. (\ref{Theta-to-zero}) to simplify the $*$-product of $\Phi$'s at the same point to a commutative (point-wise) product of $\Phi_{0}$'s. For the $*$-product of $n$ $\Phi$'s,
\beq
\label{Theta-to-zero2}
\Phi(x) * \Phi(x) * \cdots * \Phi(x) = \Big(\Phi_{0}(x)\Big)^{n} e^{\frac{1}{2}\overleftarrow{\partial}^{\mu}\theta_{\mu \nu}P^{\nu}}
\eeq
This is a very important result. Using this result, we can prove that there is no UV-IR mixing in a noncommutative field theory with matter fields and no gauge interactions \cite{Oeckl:2000eg, uv-ir}.

The interaction Hamiltonian density is built out of quantum fields. It transforms like a single scalar field in the noncommutative theory also. (This is the case only when we choose a $*$-product between the fields to write down the Hamiltonian density.) Thus a generic interaction Hamiltonian density ${\cal H}_{I}$ involving only $\Phi$'s (for simplicity) is given by
\beq
\label{Hi}
{\cal H}_{I}(x) = \Phi(x) * \Phi(x) * \cdots * \Phi(x)
\eeq
This form of the Hamiltonian and the twisted statistics of the fields is all that is required to show that there is no UV-IR mixing in this theory. This happens because the $S$-matrix becomes independent of $\theta^{\mu\nu}$.

We illustrate this result for the first nontrivial term $S^{(1)}$ in the expansion of the $S$-matrix. It is
\beq
S^{(1)} = \int d^{4}x \; {\cal H}_{I} (x).
\eeq

Using eqn. (\ref{Theta-to-zero}) we write down the interaction Hamiltonian density given in eqn. (\ref{Hi}) as
\beq
{\cal H}_{I}(x) = \Big(\Phi_{0}(x)\Big)^{n} e^{\frac{1}{2}\overleftarrow{\partial}^{\mu}\theta_{\mu \nu}P^{\nu}}
\eeq
Assuming that the fields behave ``nicely" at infinity, the integration over ~$x$~ gives
\beq
\int d^{4}x \Big(\Phi_{*}(x)\Big)^{n} = \int d^{4}x \Big(\Phi_{0}(x)\Big)^{n} e^{\frac{1}{2}\overleftarrow{\partial}^{\mu}\theta_{\mu \nu}P^{\nu}} = \int d^{4}x \Big(\Phi_{0}(x)\Big)^{n}.
\eeq

Thus $~S^{(1)}~$ is independent of $~\theta^{\mu\nu}$.
By similar calculations we can show that the $S$-operator is independent of $~\theta^{\mu \nu}~$ to all orders \cite{bal, uv-ir, bal-sasha-babar, bal-stat}.

\subsection{Covariant Derivatives of Quantum Fields}

In this section we briefly discuss how to choose appropriate covariant derivatives $D_{\mu}$ of a quantum field associated with~ ${\cal A}_{\theta}({\mathbb R}^{3+1})$.

To define the desirable properties of covariant derivatives $D_{\mu}$, let us first look at ways of multiplying the field ~$\Phi_{\theta}$~ by a function ~$\alpha_{0} \in {\cal A}_{0}({\mathbb R}^{3+1})$. There are two possibilities \cite{bal-sasha-babar}:
\bea
\Phi &\rightarrow& (\Phi_{0} \alpha_{0}) e^{\frac{1}{2}\overleftarrow{\partial}\wedge P} \equiv T_{0} (\alpha_{0})\Phi,\\
\Phi &\rightarrow& (\Phi_{0} *_{\theta} \alpha_{0}) e^{\frac{1}{2}\overleftarrow{\partial}\wedge P} \equiv T_{\theta} (\alpha_{0})\Phi
\eea
where ~$T_{0}$~ gives a representation of the commutative algebra of functions and ~$T_{\theta}$~ gives that of a $*$-algebra.

A ~$D_{\mu}$~ that can qualify as the covariant derivative of a quantum field associated with ~${\cal A}_{0}({\mathbb R}^{3+1})$~ should preserve statistics, Poincar\'e and gauge invariance and must obey the Leibnitz rule
\beq
\label{Leibnitz}
D_{\mu}(T_{0}(\alpha_{0})\Phi) = T_{0}(\alpha_{0})(D_{\mu}\Phi) + T_{0} (\partial_{\mu} \alpha_{0})\Phi
\eeq
The requirement given in eqn. (\ref{Leibnitz}) reflects the fact that $D_{\mu}$ is associated with the commutative algebra ${\cal A}_{0}({\mathbb R}^{3+1})$.

There are two immediate choices for $D_{\mu}\Phi$:
\bea
\label{firstChoice}
&&1. ~~ D_{\mu}\Phi = ((D_{\mu})_{0}\Phi_{0})e^{\frac{1}{2}\overleftarrow{\partial}\wedge P},\\
&&2. ~~ D_{\mu}\Phi = ((D_{\mu})_{0}e^{\frac{1}{2}\overleftarrow{\partial}\wedge P})(\Phi_{0})e^{\frac{1}{2}\overleftarrow{\partial}\wedge P}
\eea
where $(D_{\mu})_{0} = \partial_{\mu} + (A_{\mu})_{0}$ and $(A_{\mu})_{0}$ is the commutative gauge field, a function only of the commutative coordinates $x_{c}$.

Both the choices preserve statistics, Poincar\'e and gauge invariance, but the second choice does not satisfy eqn. (\ref{Leibnitz}). Thus we identify the correct covariant derivative in our formalism as the one given in the first choice, eqn. (\ref{firstChoice}).

\subsection{Matter fields with gauge interactions}
We assume that gauge (and gravity) fields are commutative fields, which means that they are functions only of $x^{\mu}_{c}$. For Aschieri et al. \cite{aschieri, Aschieri}, instead, they are associated with ${\cal A}_\theta(\rr^{3+1})$. Matter fields on ${\cal A}_{\theta}({\mathbb R}^{3+1})$ must be transported by the connection compatibly with  eqn.~(\ref{Theta-to-zero}), so from the previous section, we see that the natural choice for  covariant derivative is
\beq
D_{\mu} \Phi = (D_{\mu}^{c} \Phi_{0}) \; e^{\frac{i}{2}
\overleftarrow{\partial} \wedge P},
\label{eq:covariant}
\eeq
where
\beq
D_{\mu}^{c} \Phi_{0} = \partial_{\mu}\Phi_{0} + A_{\mu}\Phi_{0}\; ,
\eeq
$P_{\mu}$ is the total momentum operator for all the fields and the fields $A_{\mu}$ and $\Phi_{0}$ are multiplied point-wise,
\beq
A_{\mu}\Phi_{0}(x)=A_{\mu}(x)\Phi_{0}(x).
\eeq

Having identified the correct covariant derivative, it is simple to write down the Hamiltonian for gauge theories. The commutator of two covariant derivatives gives us the curvature. On using eqn. (\ref{eq:covariant}),
\begin{eqnarray}
[D_{\mu}, D_{\nu}] \Phi &=& \Big([D^{c}_{\mu}, D^{c}_{\nu}]\Phi_{0}\Big)e^{\frac{i}{2}\overleftarrow{\partial} \wedge P} \\
&=&\Big(F_{\mu \nu}^{c}\Phi_{0}\Big)e^{\frac{i}{2}\overleftarrow{\partial} \wedge
P}.
\end{eqnarray}
As $F_{\mu \nu}^{c}$ is the standard $\theta^{\mu \nu}=0$ curvature, our gauge field
is associated with ${\cal A}_{0}({\mathbb R}^{3+1})$. Thus pure gauge theories on the GM plane are identical to their counterparts on commutative spacetime. (For Aschieri et al. \cite{aschieri} the
curvature would be the $\star$-commutator of $D_{\mu}$'s.)

The gauge theory formulation we adopt here is fully explained in \cite{bal-sasha-babar}. It differs from the formulation of Aschieri et al. \cite{aschieri} (where covariant derivative is defined using star product) and has the advantage of being able to accommodate any gauge group and not just $U(N)$ gauge groups and their direct products. The gauge theory formulation we adopt here thus avoids multiplicity of fields that the expression for covariant derivatives with $\star$ product entails.

In the single-particle sector (obtained by taking the matrix element of eqn.~(\ref{eq:covariant}) between vacuum and one-particle states), the $P$ term can be dropped and we get for a single particle wave function $f$ of a particle associated with $\Phi$,
\beq
D_{\mu}f(x) = \partial_{\mu}f(x)+A_{\mu}(x)f(x).
\eeq
Note that we can also write $D_{\mu}\Phi$ using $\star$-product:
\beq
D_{\mu}\Phi = \Big(D_{\mu}^{c} e^{\frac{i}{2} \overleftarrow{\partial} \wedge
P}\Big)\star \Big(\Phi_{0}e^{\frac{i}{2} \overleftarrow{\partial} \wedge
P}\Big).
\eeq
Our choice of covariant derivative allows us to write the interaction Hamiltonian density
for pure gauge fields as follows:
\beq
{\cal H}_{I \theta}^{^G} = {\cal H}_{I 0}^{^G}.
\eeq

For a theory with matter and gauge fields, the interaction Hamiltonian density splits into two parts,
\beq
{\cal H}_{I \theta} = {\cal H}^{^{M, G}}_{I \theta}+{\cal H}^{^G}_{I \theta},
\eeq
where
\bea
{\cal H}^{^{M, G}}_{I \theta}&=&{\cal H}^{^{M, G}}_{I 0} \; e^{\frac{i}{2}
\overleftarrow{\partial} \wedge P},\nn \\
{\cal H}^{^G}_{I \theta}&=&{\cal H}^{^G}_{I 0}.
\eea
The matter-gauge field couplings are also included in ${\cal H}^{^{M, G}}_{I \theta}$.

In quantum electrodynamics ($QED$), ${\cal H}^{^G}_{I \theta}=0$. Thus the $S$-operator for the twisted $QED$ is the same for the untwisted $QED$:
\beq
S^{^{QED}}_{\theta}=S^{^{QED}}_{0}.
\eeq

\begin{figure}
\centerline{\epsfig{figure=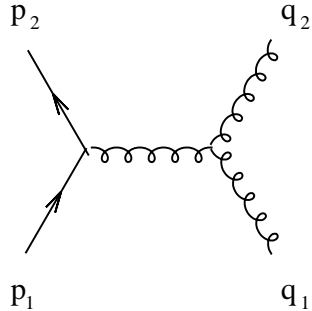,clip=4cm,width=4cm}}
\caption{A Feynman diagram in QCD with non-trivial $\theta$-dependence. The twist of ${\cal H}^{^{M, G}}_{I 0}$ changes the gluon propagator. The propagator is different from the usual one by its dependence on terms of the form $\vec{\theta}^{0} \cdot {\bf P}_{in}$, where $(\vec{\theta}^{0})_{i} = \theta^{0i}$ and ${\bf P}_{in}$ is the total momentum of the incoming particles. Such a frame-dependent modification violates Lorentz invariance.}
\label{fig:qcd}
\end{figure}

In a non-abelian gauge theory, ${\cal H}^{^G}_{\theta}={\cal H}^{^G}_{0} \neq 0$, so that in the presence of nonsinglet matter fields \cite{bal-sasha-babar},
\beq
S^{^{M, G}}_{\theta} \neq S^{^{M, G}}_{0},
\eeq
because of the cross-terms between ${\cal H}^{^{M, G}}_{I \theta}$ and ${\cal H}^{^{G}}_{I \theta}$. In particular, this inequality happens in QCD. One such example is the quark-gluon scattering through a gluon exchange. The Feynman diagram for this process is given in Fig. \ref{fig:qcd}.

\subsection{Causality and Lorentz Invariance}
The very process of replacing the point-wise multiplication of functions at the same point by a $*$-multiplication makes the theory non-local. The $*$-product contains an infinite number of space-time derivatives and this in turn affects the fundamental causal structure on which all local, point-like quantum field theories are built upon.

Let ${\cal H}_{I}$ be the interaction Hamiltonian density in the interaction representation. The interaction representation $S$-matrix is
\beq
S = \textrm{T exp} \Big(-i \int d^{4}x \; {\cal H}_{I}(x)\Big).
\eeq

In a commutative theory, the interaction Hamiltonian density ${\cal H}_{I}$ satisfies the Bogoliubov - Shirkov \cite{Bogoliubov} causality
\beq
\label{causality1}
[{\cal H}_{I}(x), {\cal H}_{I}(y)] = 0, \; \; \; x \sim y
\eeq
where $x \sim y$ means $x$ and $y$ are space-like separated.

This causality relation plays a crucial role in maintaining the Lorentz invariance in all the local, point-like quantum field theories. Weinberg \cite{Weinberg1, Weinberg2} has discussed the fundamental significance of this equation in connection with the relativistic invariance of the $S$-matrix. If eqn. (\ref{causality1}) fails, $S$ cannot be relativistically invariant.

To see why this is the case, we consider the lowest term ~$S^{(2)}$~ of the ~$S$-matrix containing non-trivial time ordering. It is ~$S^{(2)}=-\frac{1}{2}\int d^4xd^4y~T(~{\cal H}_{I}(x){\cal H}_{I}(y)~),$ where
\begin{eqnarray}
T(~{\cal H}_{I}(x){\cal H}_{I}(y)~)&:=&\theta(x^0-y^0){\cal H}_{I}(x){\cal H}_{I}(y) +
\theta(y^0-x^0){\cal H}_{I}(y){\cal H}_{I}(x)\nonumber\\
&=& {\cal H}_{I}(x){\cal H}_{I}(y)+ ( \theta(x^0-y^0)-1){\cal H}_{I}(x){\cal H}_{I}(y)+ \theta(y^0-x^0){\cal H}_{I}(y){\cal H}_{I}(x)\nonumber\\
&=& {\cal H}_{I}(x){\cal H}_{I}(y)-\theta(y^0-x^0)[{\cal H}_{I}(x),{\cal H}_{I}(y)]. 
\end{eqnarray}

If $U(\Lambda)$ is the unitary operator  on the quantum Hilbert space for implementing the Lorentz transformation ~$\Lambda$~ connected to the identity, that is, ~$\Lambda\in {P}^\uparrow_+$, then

$$U(\Lambda)T( {\cal H}_I(x){\cal H}_I(y) )U(\Lambda)^{-1}={\cal H}_I(\Lambda x){\cal H}_I(\Lambda y)-\theta(y^0-x^0)[{\cal H}_I(\Lambda x),{\cal H}_I(\Lambda y)].$$
If this is equal to~ $T( {\cal H}_I(\Lambda x){\cal H}_I(\Lambda y) )$, that is, if

$$  \theta( y^0- x^0 )[{\cal H}_I(\Lambda x),{\cal H}_I(\Lambda y)]= \theta((\Lambda y)^0-(\Lambda x)^0 )[( {\cal H}_I(\Lambda x),{\cal H}_I(\Lambda y) ],$$
then ~$S^{(2)}$~ is invariant under ~$\Lambda\in {P}^\uparrow_+$. It is clearly invariant under translations. Hence the invariance of $S^{(2)}$ under $ {P}^\uparrow_+$ requires that either  ~$\theta(y^0-x^0)$~ is invariant or that ~$[{\cal H}_{I}(x),{\cal H}_{I}(y)]=0$.

When  ~$x\nsim y$, the time step function ~$\theta(y^0-x^0)$~ is invariant under  ~$ P_{+}^{\uparrow}$~ since ~$\Lambda\in  {P}^\uparrow_+$~ cannot reverse the direction of time.

However, when ~$x\sim y$, ~$\Lambda\in  {P}^\uparrow_+$~ can reverse the direction of time and so ~$\theta(y^0-x^0)$~ is not invariant. One therefore requires that ~$[{\cal H}_{I}(x),{\cal H}_{I}(y)]=0$ if $x\sim y$. Therefore a commonly imposed condition for the invariance of ~$S^{(2)}$~ under ~${P}^\uparrow_+$~ is

\beq \label{causality2}[({\cal H}_{I}(x),{\cal H}_{I}(y))]=0 ~~~~\text{whenever}~~~~ x\sim y.\eeq

One can show by similar arguments that it is natural to impose the causality condition $(\ref{causality2})$ to maintain the ~${P}^\uparrow_+$~ invariance of of the general term

$$S^{(n)}=\frac{(-i)^n}{n!}\int d^4x_1d^4x_2...d^4x_n~T(~{\cal H}_{I}(x_1){\cal H}_{I}(x_2)...{\cal H}_{I}(x_n)~),$$
in $S$.  Here

\begin{eqnarray}
 &&T(~{\cal H}_{I}(x_1){\cal H}_{I}(x_2)...{\cal H}_{I}(x_n)~)\nonumber\\
 &&~~~~~=\sum_{i_1,...,i_n~\in
\{1,2,...,n\}}~\theta(x_{i_1}-x_{i_2})\theta(x_{i_2}-x_{i_3})...\theta(x_{i_{n-1}}-x_{i_{n}})~{\cal H}_{I}(x_{i_1}){\cal H}_{I}(x_{i_2})...{\cal H}_{I}(x_{i_{n}});\nonumber
\end{eqnarray}
where ~~$i_j\neq i_k$~~ if ~~$j\neq k$. 

In a noncommutative theory, due to twisted statistics, the interaction Hamiltonian density might not satisfy (\ref{causality2}) but $S$ can still be Lorentz-invariant. For example, consider the interaction Hamiltonian density for the electron-photon system
\beq
{\cal H}_{I}(x) = i e \; (\bar{\psi} \star \gamma^{\rho}A_{\rho}\psi)(x).
\eeq

For simplicity, we consider the case where $\theta^{0i}=0$ and $\theta^{ij} \neq 0$. We write down the $S$-matrix
\bea
S &=& \textrm{T exp} \Big(-i \int d^{3}x {\cal H}_{I}(x)\Big)
\eea
where ${\cal H}_{I}(x) = i e \; (\bar{\psi}\gamma^{\rho}A_{\rho}\psi)(x).$ Here we have used the property of the Moyal product to remove the $*$ in ${\cal H}_{I}$ while integrating over the spatial variables. The fields $\psi$ and $\bar{\psi}$ are still noncommutative as their oscillator modes contain $\theta^{\mu \nu}$.

We can write down ${\cal H}_{I}(x)$ in the form
\beq
{\cal H}_{I}(x) ={\cal H}^{(0)}_I(x)e^{\frac{1}{2}\overleftarrow{\partial}\wedge\overrightarrow{P} }
\eeq
where ${\cal H}^{(0)}_I$ gives the interaction Hamiltonian for $\theta^{\mu\nu}=0$ and satisfies the causality condition $(\ref{causality2})$. It follows that ${\cal H}_I$ does not fulfill the causality condition $(\ref{causality2})$. Still, as shown in $\cite{bal-sasha-babar}$, $S$ is Lorentz invariant. (For further discussion, see $\cite{bal-sasha-babar}$.)
\section{Discrete Symmetries - ${\bf C}$, ${\bf P}$, ${\bf T}$ and ${\bf CPT}$}
So far our discussion was centered around the identity component $P_{+}^{\uparrow}$ of the Lorentz group $P$. In this section we investigate the symmetries of our noncommutative theory under the action of discrete symmetries - parity ${\bf P}$, time reversal ${\bf T}$, charge conjugation ${\bf C}$ and their combined operation ${\bf CPT}$. The ${\bf CPT}$ theorem \cite{pct, cpt} is very fundamental in nature and all local relativistic quantum field theories are ${\bf CPT}$ invariant. Quantum field theories on the GM plane are non-local and so it is important to investigate the validity of the ${\bf CPT}$
theorem in these theories.

\subsection{Transformation of Quantum Fields Under ${\bf C}$, ${\bf P}$ and ${\bf T}$ }
Under \textbf{C}, the Poincar\'e  group ~$P_+^{\uparrow}$, the creation and annihilation operators ~$c_{\bf k}$, $c^\dagger_{\bf k}$, $d_{\bf k}$, $d^{\dagger}_{\bf k}$ of a second quantized field transform in the same way as their counterparts in an untwisted theory \cite{bal-sasha-babar}. Using the dressing transformation $\cite{Grosse, Faddeev-Zamolodchikov}$, we can then deduce the transformation laws for ~$a_\textbf{k}, ~a^\dagger_\textbf{k},~ b_\textbf{k},~ b^\dagger_\textbf{k}$, and the quantum fields. They automatically imply the appropriate twisted coproduct in the matter sector (and of course the untwisted coproduct for gauge fields.) It then implies the transformation laws for the fields under the full group generated by ${\bf C}$ and ${\cal P}$ by the group properties of that group: they are all induced from those of $c_{\bf k}$, $c^\dagger_{\bf k}$, $d_{\bf k}$, $d^{\dagger}_{\bf k}$ in the above fashion. (We always try to preserve such group properties.) We make use of this observation when we discuss the transformation properties of quantum fields under discrete symmetries.

So far we have not mentioned the transformaton property of the noncommutativity parameter $\theta^{\mu \nu}$. The matrix $\theta^{\mu \nu}$ is a constant antisymmetric matrix. In the approach using the twisted coproduct for the Poincar\'e group, $\theta^{\mu \nu}$ is {\it not} transformed by Poincar\'e transformations or in fact by any other symmetry: they are truly constants. Nevertheless Poincar\'e invariance and other symmetries can be certainly recovered for interactions invariant under the twisted symmetry actions at the level of classical theory and also for Wightman functions \cite{drinfeld, bal-stat, aschieri, dimitrijevic}.

We discuss the transformation of quantum fields under the action of discrete symmetries below.

\subsubsection{Charge conjugation ${\bf C}$}
The charge conjugation operator is not a part of the Lorentz group and commutes with $P_{\mu}$ (and in fact with the full Poincar\'e group). This implies that the coproduct \cite{chaichian, aschieri} for the charge conjugation operator ${\bf C}$ in the twisted case is the same as the coproduct for ${\bf C}$ in the untwisted case. So, we write
\beq
\label{coproduct-for-C}\Delta_{\theta}({\bf C}) = \Delta_{0}({\bf C}) = {\bf C} \otimes {\bf C},
\eeq
with the understanding that \textbf{C} is an element of the group algebra $ G^\ast$, where $G=\{\textrm{\textbf{C}}\}\times P^\uparrow_+$. (This is why we use $\otimes$ and not $\times$  in (\ref{coproduct-for-C}).)

Under charge conjugation,
\beq
c_{{\bf k}} \stackrel{{\bf C}}\longrightarrow d_{{\bf k}}, \; \; \; a_{{\bf k}} \stackrel{{\bf C}}\longrightarrow b_{{\bf k}}
\eeq
where the twisted operators are related to the untwisted ones by the dressing transformation \cite{Grosse, Faddeev-Zamolodchikov}: ~$a_{{\bf k}}=c_{{\bf k}} \; e^{-\frac{i}{2} k \wedge P}$ and  ~$b_{{\bf k}}=d_{{\bf k}} \; e^{-\frac{i}{2} k \wedge P}$.

It follows that
\beq
\varphi_{\theta} \stackrel{{\bf C}}\longrightarrow \varphi^{_{\bf C}}_{0}\; e^{\frac{1}{2} \overleftarrow{\partial} \wedge P}, \; \; \varphi_{0}^{_{\bf C}} = {\bf C} \varphi_{0} {\bf C}^{-1}.
\eeq
while the $\ast$-product of two such fields $\varphi_\theta$ and $\chi_\theta$ transforms according to
\bea
\varphi_{\theta} \star \chi_{\theta} &=&(\varphi_{0} \chi_{0})\; e^{\frac{1}{2} \overleftarrow{\partial} \wedge P} \nn \\ &\stackrel{{\bf C}}\longrightarrow& ({\bf C} \varphi_{0}\chi_{0}{\bf C}^{-1})\; e^{\frac{1}{2} \overleftarrow{\partial} \wedge P} \nn \\
&=&(\varphi_{0}^{{\bf C}} \chi_{0}^{{\bf C}})\; e^{\frac{1}{2}\overleftarrow{\partial} \wedge P}.
\eea

\subsubsection{Parity ${\bf P}$}
Parity is a unitary operator on ${\cal A}_{0}({\mathbb R}^{3+1})$. But parity transformations do not induce automorphisms of ${\cal A}_{\theta}({\mathbb R}^{3+1})$ \cite{bal-unitary} if its coproduct is
\beq
\Delta_{0}({\bf P})={\bf P} \otimes {\bf P}.
\eeq
That is, this coproduct is not compatible with the $\star$-product. Hence the coproduct for parity is not the same as that for the $\theta^{\mu \nu}=0$ case.

But the twisted coproduct $\Delta_{\theta}$, where
\beq
\Delta_\theta({\bf P}) = {\cal F}_{\theta}^{-1} \; \Delta_{0} ({\bf P}) \; {\cal
F}_{\theta},
\eeq
{\it is} compatible with the $\star$-product. So, for ${\bf P}$ as well, compatibility with the $\star$-product fixes the coproduct \cite{bal}.

Under parity,
\beq
c_{{\bf k}} \stackrel{{\bf P}}\longrightarrow c_{-{\bf k}}, \; \; \; \; d_{{\bf k}}
\stackrel{{\bf P}}\longrightarrow d_{-{\bf k}}
\label{eq:cdP}
\eeq
and hence
\beq
a_{{\bf k}} \stackrel{{\bf P}}\longrightarrow a_{-{\bf k}} \;
e^{i(k_{0}\theta^{0i}P_{i}-k_{i}\theta^{i0}P_{0})}, \; \; \; \; b_{{\bf k}}
\stackrel{{\bf P}}\longrightarrow b_{-{\bf k}} \;
e^{i(k_{0}\theta^{0i}P_{i}-k_{i}\theta^{i0}P_{0})}.
\label{eq:abP}
\eeq
By an earlier remark \cite{bal-sasha-babar}, eqns. (\ref{eq:cdP}) and (\ref{eq:abP}) imply the transformation law for twisted scalar fields. A twisted complex scalar field $\varphi_{\theta}$ transforms under parity as follows,
\bea
\varphi_{\theta}&=& \varphi_{0} \; e^{\frac{1}{2}\overleftarrow{\partial} \wedge P}~\stackrel{{\bf P}}\longrightarrow~ {\bf P}\Big(\varphi_{0} \; e^{\frac{1}{2}\overleftarrow{\partial} \wedge P}\Big){\bf P}^{-1}
~=~\varphi_{0}^{_{\bf P}}\; e^{\frac{1}{2} \overleftarrow{\partial} \wedge (P_{0}, -\overrightarrow{P})},
\eea
where ~$\varphi_{0}^{_{\bf P}} = {\bf P} \varphi_{0} {\bf P}^{-1}$~ and ~$\overleftarrow{\partial} \wedge (P_{0}, -\overrightarrow{P}) := -\overleftarrow{\partial}_{0}
\theta^{0i} P_{i} -\overleftarrow{\partial}_{i} \theta^{ij}P_{j}+ \overleftarrow{\partial}_{i} \theta^{i0}P_{0}$.

The product of two such fields $\varphi_{\theta}$ and $\chi_{\theta}$ transforms according to
\bea
\varphi_{\theta} \star \chi_{\theta} &=&(\varphi_{0} \chi_{0})\; e^{\frac{1}{2} \overleftarrow{\partial} \wedge P}\stackrel{{\bf P}}\longrightarrow (\varphi^{_{\bf P}}_{0} \chi^{_{\bf P}}_{0})\; e^{\frac{1}{2} \overleftarrow{\partial} \wedge (P_{0}, -\overrightarrow{P})}
\eea

Thus fields transform under ${\bf P}$ with an extra factor $e^{-(\overleftarrow{\partial}_{0}\theta^{0i}P_{i} + \partial_{i}\theta^{ij}P_{j})} = e^{-\overleftarrow{\partial}_{\mu}\theta^{\mu j}P_{j}}$ when $\theta^{\mu \nu} \neq 0$.

\subsubsection{Time reversal ${\bf T}$}
Time reversal ~${\bf T}$~ is an anti-linear operator. Due to antilinearity, ~${\bf T}$~ induces automorphisms on ~${\cal A}_{\theta}({\mathbb R}^{3+1})$~ for the coproduct

$$\Delta_0(T)=T\otimes T~~~~\textrm{if}~~\theta^{ij}=0,$$
but not otherwise.

Under time reversal,
\beq
c_{{\bf k}} \stackrel{{\bf T}}\longrightarrow c_{-{\bf k}}, \; \; \; \; d_{{\bf k}}
\stackrel{{\bf T}}\longrightarrow d_{-{\bf k}}
\eeq
\beq
a_{{\bf k}} \stackrel{{\bf T}}\longrightarrow a_{-{\bf k}} \;
e^{-i(k_{i}\theta^{ij}P_{j})}, \; \; \; \; b_{{\bf k}} \stackrel{{\bf T}}\longrightarrow
b_{-{\bf k}} \; e^{-i(k_{i}\theta^{ij}P_{j})}.
\eeq

When ~$\theta^{\mu\nu}\neq 0$,~ compatibility with the $\star$-product fixes the coproduct for ~${\bf T}$~ to be
\beq
\Delta_\theta({\bf T}) = {\cal F}_{\theta}^{-1} \; \Delta_{0} ({\bf T}) \; {\cal F}_{\theta}.
\eeq

This coproduct is also required in order to maintain the group properties of ${\cal P}$,
the full Poincar\'e group.

A twisted complex scalar field $\varphi_{\theta}$ hence transforms under time reversal as
follows,
\bea
\varphi_{\theta}&=& \varphi_{0} \; e^{\frac{1}{2}\overleftarrow{\partial} \wedge P}~~\stackrel{{\bf T}}\longrightarrow ~~\varphi_{0}^{_{\bf T}}\; e^{\frac{1}{2} \overleftarrow{\partial} \wedge (P_{0}, -\overrightarrow{P})},
\eea
where $\varphi_0^T=T\varphi_0T^{-1}$,
while the product of two such fields $\varphi_{\theta}$ and $\chi_{\theta}$ transforms according to
\bea
\varphi_{\theta} \star \chi_{\theta} &=&(\varphi_{0} \chi_{0})\; e^{\frac{1}{2} \overleftarrow{\partial} \wedge P}~~\stackrel{{\bf T}}\longrightarrow~~ (\varphi^{_{\bf T}}_{0} \chi^{_{\bf T}}_{0})\; e^{\frac{1}{2} \overleftarrow{\partial} \wedge (P_{0}, -\overrightarrow{P})}
\eea

Thus the time reversal operation as well induces an extra factor $e^{-\overleftarrow{\partial}_{i}\theta^{ij}P_{j}}$ in the transformation
property of fields when $\theta^{\mu \nu} \neq 0$.

\subsubsection{${\bf CPT}$}
When ${\bf CPT}$ is applied,
\beq
c_{{\bf k}} \stackrel{{\bf CPT}}\longrightarrow d_{{\bf k}}, \; \; \; \; d_{{\bf k}}
\stackrel{{\bf CPT}}\longrightarrow c_{{\bf k}},
\eeq
\beq
a_{{\bf k}} \stackrel{{\bf CPT}}\longrightarrow b_{{\bf k}}e^{i(k \wedge P)}, \; \; \; \;
b_{{\bf k}} \stackrel{{\bf CPT}}\longrightarrow a_{{\bf k}}e^{i(k \wedge P)}.
\eeq

The coproduct for ${\bf CPT}$ is of course
\beq
\Delta_\theta({\bf CPT}) = {\cal F}_{\theta}^{-1} \; \Delta_{0} ({\bf CPT}) \; {\cal F}_{\theta}.
\eeq

A twisted complex scalar field $\varphi_{\theta}$ transforms under ${\bf CPT}$ as follows,
\bea
\varphi_{\theta}&=& \varphi_{0} \; e^{\frac{1}{2}\overleftarrow{\partial} \wedge P}\nn \\
&\stackrel{{\bf CPT}}\longrightarrow& {\bf CPT}\Big(\varphi_{0} \; e^{\frac{1}{2}\overleftarrow{\partial} \wedge P}\Big) ({\bf CPT})^{-1}\nn \\
&=& \varphi_{0}^{_{{\bf CPT}}}\; e^{\frac{1}{2} \overleftarrow{\partial} \wedge P},
\eea
while the product of two such fields $\varphi_{\theta}$ and $\chi_{\theta}$ transforms according to
\bea
\varphi_{\theta} \star \chi_{\theta} &=&(\varphi_{0} \chi_{0})\; e^{\frac{1}{2} \overleftarrow{\partial} \wedge P} \nn \\ &\stackrel{{\bf CPT}}\longrightarrow& (\varphi^{_{{\bf CPT}}}_{0} \chi^{_{{\bf CPT}}}_{0})\; e^{\frac{1}{2} \overleftarrow{\partial} \wedge P}.
\eea
\subsection{{\bf CPT} in  Non-Abelian Gauge Theories}
The standard model, a non-abelian gauge theory, is ${\bf CPT}$ invariant, but it is not invariant under ${\bf C}$, ${\bf P}$, ${\bf T}$ or products of any two of them. So we focus on discussing just ${\bf CPT}$ for its $S$-matrix when $\theta^{\mu \nu} \neq 0$. The discussion here can be easily adapted to any other non-abelian gauge theory.

\subsubsection{Matter fields coupled to gauge fields}
The interaction representation $S$-matrix is
\beq
{S}^{^{M, G}}_{\theta} = \text{T exp} \; \Big[{-i\int d^{4}x \; {\cal H}^{^{M, G}}_{I
\theta}(x)}\Big]
\eeq
where ${\cal H}^{^{M, G}}_{I \theta}$ is the interaction Hamiltonian density for matter
fields (including also matter-gauge field couplings). Under ${\bf CPT}$,
\beq
{\cal H}^{^{M, G}}_{I \theta}(x) \stackrel{{\bf CPT}}\longrightarrow {\cal H}^{^{M, G}}_{I
\theta}(-x)e^{\overleftarrow{\partial} \wedge P}
\eeq
where $\overleftarrow{\partial}$ has components $\frac{\overleftarrow{\partial}}{\partial x_{\mu}}$. We write ${\cal H}^{^{M, G}}_{I \theta}$ as
\beq
{\cal H}^{^{M, G}}_{I \theta} = {\cal H}^{^{M, G}}_{I 0} \; e^{\frac{1}{2}
\overleftarrow{\partial} \wedge P}.
\label{eq:matter}
\eeq
Thus we can write the interaction Hamiltonian density after ${\bf CPT}$ transformation in
terms of the untwisted interaction Hamiltonian density:
\bea
{\cal H}^{^{M, G}}_{I \theta}(x) \; \; \stackrel{{\bf CPT}}\longrightarrow&& {\cal H}^{^{M, G}}_{I
\theta}(-x)\; e^{\overleftarrow{\partial} \wedge P} \nn \\ &=& {\cal
H}^{^{M, G}}_{I0}(-x)\; e^{-\frac{1}{2}\overleftarrow{\partial} \wedge
P}\; e^{\overleftarrow{\partial} \wedge P}\nn \\
&=&{\cal H}^{^{M, G}}_{I0}(-x)\; e^{\frac{1}{2}\overleftarrow{\partial} \wedge P}.
\eea

Hence under ${\bf CPT}$,
\beq
{S}^{^{M, G}}_{\theta} = \text{T exp} \; \Big[-i\int d^{4}x \; {\cal H}^{^{M, G}}_{I
0}(x) \; e^{\frac{1}{2}\overleftarrow{\partial}\wedge P}\Big] \rightarrow \text{T exp} \; \Big[i\int d^{4}x \; {\cal H}^{^{M, G}}_{I
0}(x) \; e^{-\frac{1}{2}\overleftarrow{\partial}\wedge P}\Big] = ({S}^{^{M, G}}_{-\theta})^{-1}. \nn
\eeq

But it has been shown elsewhere that ${S}^{^{M, G}}_{\theta}$ is independent of $\theta$ \cite{uv-ir}. Hence also ${S}^{^{M, G}}_{\theta}$ is independent of $\theta$.

Therefore a quantum field theory with no pure gauge interaction is ${\bf CPT}$ ``invariant" on ${\calA}_{\theta}({\mathbb R}^{3+1})$. In particular quantum electrodynamics ($QED$) preserves ${\bf CPT}$.

\subsubsection{Pure Gauge Fields}
The interaction Hamiltonian density for pure gauge fields is independent of $\theta^{\mu \nu}$ in the approach of \cite{bal-sasha-babar}:
\beq
{\cal H}_{I \theta}^{^G} = {\cal H}_{I 0}^{^G}\; .
\eeq

Hence also the $S$ becomes $\theta$-independent,
\beq
{S}^{^G}_{\theta} = {S}^{^G}_{0},
\eeq

and ${\bf CPT}$ holds as a good ``symmetry" of the theory.

\subsubsection{Matter and Gauge Fields}

All interactions of matter and gauge fields can be fully discussed by writing the $S$-operator as\beq
{{\bf S}}^{^{M,G}}_{\theta} = \text{T exp} \; \Big[{-i\int d^{4}x \; {\cal
H}_{I
\theta}(x)}\Big],
\eeq
\beq
{\cal H}_{I \theta} = {\cal H}^{^{M, G}}_{I \theta}+{\cal H}^{^G}_{I \theta},
\eeq
where
\beq
{\cal H}^{^{M, G}}_{I \theta}={\cal H}^{^{M, G}}_{I 0} \; e^{\frac{1}{2}
\overleftarrow{\partial} \wedge P}\nn
\eeq
and
\beq
{\cal H}^{^G}_{I \theta}={\cal H}^{^G}_{I 0}\; .\nn
\eeq

In $QED$, ${\cal H}^{^G}_{I \theta}=0$. Thus the $S$-operator ${\bf S}^{^{QED}}_{\theta}$ is
the same as for the $\theta^{\mu \nu} =0$. That is,
\beq
{\bf S}^{^{QED}}_{\theta}={\bf S}^{^{QED}}_{0}.
\eeq

Hence ${\bf C}$, ${\bf P}$, ${\bf T}$ and ${\bf CPT}$ are good ``symmetries" for $QED$ on the GM plane.

For a non-abelian gauge theory with non-singlet matter fields, ${\cal H}^{^G}_{I \theta}={\cal H}^{^G}_{I 0} \neq 0$ so that if ${\bf S}^{^{M, G}}_{\theta}$ is the $S$-matrix of the theory,
\beq
{\bf S}^{^{M, G}}_{\theta} \neq {\bf S}^{^{M, G}}_{0}.
\eeq

The $S$-operator ${\bf S}^{^{M,G}}_{\theta}$ depends only on $\theta^{0i}$ in a
non-abelian theory, that is,  ${\bf S}^{^{M,G}}_{\theta} =
{\bf S}^{^{M,G}}_{\theta}|_{\theta^{ij}=0}$. Applying ${\bf C}$, ${\bf P}$ and ${\bf T}$ on ${\bf S}^{^{M,G}}_{\theta}$ we can see that ${\bf C}$ and ${\bf T}$ do not affect $\theta^{0i}$ while ${\bf P}$ changes its sign. Thus a non-zero $\theta^{0i}$ contributes to ${\bf P}$ and ${\bf CPT}$ violation.

\subsection{On Feynman Graphs}

This section uses the results of \cite{bal-sasha-babar} and \cite{bal-sasha-queiroz} where Feynman rules are fully developed and field theories are analyzed further.

\begin{figure}
\centerline{\epsfig{figure=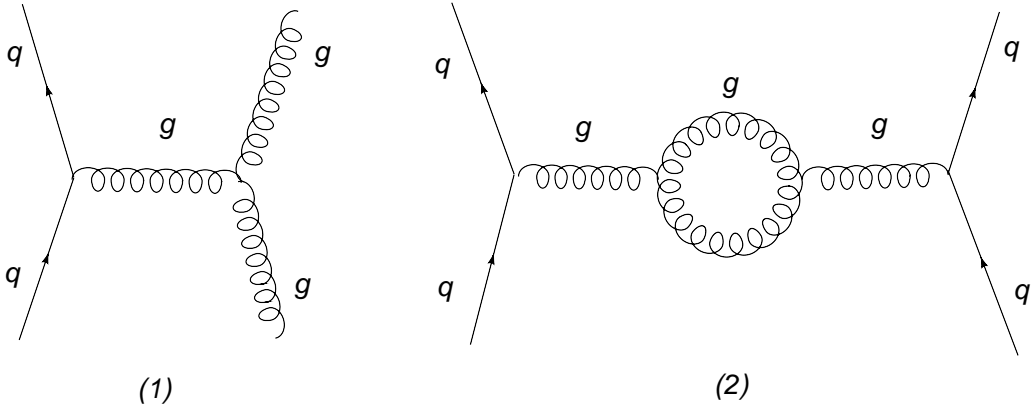}}
\caption{${\bf CPT}$ violating processes on GM plane. (1) shows quark-gluon scattering with a three-gluon vertex. (2) shows a gluon-loop contribution to quark-quark scattering.}
\label{cpt}
\end{figure}

In non-abelian gauge theories, ${\cal H}^{^{G}}_{I \theta}={\cal H}^{^{G}}_{I 0}$ is not zero as gauge fields have self-interactions. The preceding discussions show that the effects of $\theta^{\mu \nu}$ can show up only in Feynman diagrams which are sensitive to products of ${\cal H}^{^{M, G}}_{I \theta}$'s with ${\cal H}^{^{G}}_{I 0}$'s. Fig. (\ref{cpt}) shows two such diagrams.

As an example, consider the first diagram in Fig. (\ref{cpt}) To lowest order, it depends on $\theta^{0i}$.

We can substitute eqn. (\ref{eq:matter}) for ${\cal H}^{^{M, G}}_{I \theta}$ and integrate over ${\bf x}$. That gives,
\beq
{\bf S}^{(2)} =-\frac{1}{2} \int d^{4}x d^{4}y \;  \textrm{T} \Big({\cal H}_{I 0}^{^{M, G}}(x)\; e^{\frac{1}{2} \overleftarrow{\partial}_{0} \theta^{0i} P_{i}}{\cal H}_{I 0}^{^G}(y)\Big)\nn
\eeq
where $\overleftarrow{\partial}_{0}$ acts {\it only} on ${\cal H}_{I 0}^{^{M, G}}(x)$ (and not on the step functions in time entering in the definition of $\textrm{T}$.)

Now $P_{i}$, being components of spatial momentum, commutes with
\beq
\int d^{3}y \; {\cal H}^{^{G}}_{I 0}(y)\nn
\eeq
and hence for computing the matrix element defining the process ({\it 1}) in Fig. (\ref{cpt}), we can substitute $\overrightarrow{P}_{\textrm{in}}$ for $\overrightarrow{P}$,  $\overrightarrow{P}_{\textrm{in}}$ being the total incident spatial momentum:
\beq
{\bf S}^{(2)} =-\frac{1}{2} \int d^{4}x d^{4}y \;  \textrm{T} \Big({\cal H}_{I 0}^{^{M, G}}(x)\; e^{\frac{1}{2} \overleftarrow{\partial}_{0} \theta^{0i} P^{\textrm{in}}_{i}}{\cal H}_{I 0}^{^G}(y)\Big).
\eeq

Thus ${\bf S}^{(2)}$ depends on $\theta^{0i}$ unless
\beq
\theta^{0i}P^{\textrm{in}}_{i} = 0.
\eeq

This will happen in the center-of-mass system or more generally if $\overrightarrow{\theta^{0}} = $($\theta^{01}$, $\theta^{02}$, $\theta^{03}$) is perpendicular to $\overrightarrow{P}^{\textrm{in}}$.

Under ${\bf P}$ and ${\bf CPT}$, $\theta^{0i} \rightarrow -\theta^{0i}$. This shows clearly that in a general frame, $\theta^{0i}$ contributes to ${\bf P}$ violation and causes ${\bf CPT}$ violation.

The dependence of $S^{(2)}$ on the incident total spatial momentum shows that the scattering matrix is not Lorentz invariant. This noninvariance is caused by the nonlocality of the interaction Hamiltonian density: if we evaluate it at two spacelike separated points, the resultant operators do not commute. Such a violation of causality can lead to Lorentz-noninvariant $S$-operators \cite{bal-sasha-babar}.

The reasoning which reduced $e^{\frac{1}{2}\overleftarrow{\partial} \wedge P}$ to $e^{\frac{1}{2}\overleftarrow{\partial}_{0} \theta^{0i} P^{\textrm{in}}_{i}}$ is valid to all such factors in an arbitrary order in the perturbation expansion of the $S$-matrix and for arbitrary processes, $\overrightarrow{P}^{\textrm{in}}$ being the total incident spatial momentum. As $\theta^{\mu \nu}$ occur only in such factors, this leads to an interesting conclusion: if scattering happens in the
center-of-mass frame, or any frame where $\theta^{0i}P^{\textrm{in}}_{i} = 0$, then the
$\theta$-dependence goes away from the $S$-matrix. That is, $P$ and $CPT$ remain intact if $\theta^{0i}P^{\textrm{in}}_{i} = 0$. The theory becomes $P$ and $CPT$ violating in all other frames.

Terms with products of ${\cal H}^{^{M, G}}_{I \theta}$ and ${\cal H}^{^G}_{I \theta}$ are
$\theta$-dependent and they violate ${\bf CPT}$. Electro-weak and $QCD$ processes will thus
acquire dependence on $\theta$. This is the case when a diagram involves products of
${\cal H}^{^{M, G}}_{I \theta}$ and ${\cal H}^{^G}_{I \theta}$. For example
quark-gluon and quark-quark scattering on the GM plane become $\theta$-dependent ${\bf CPT}$
violating processes (See Fig. (\ref{cpt})).

These effects can be tested experimentally.
\section{Acknowledgements}
This work was partially supported by the US Department of Energy under grant number DE-FG02-85ER40231. A. P. B. warmly thanks Prof. Viqar Husain for the wonderful hospitality he enjoyed at Fredericton.

\end{document}